\documentclass[fleqn,usenatbib]{mnras}

\usepackage{newtxtext,newtxmath}

\usepackage[T1]{fontenc}
\usepackage{ae,aecompl}

\usepackage{graphicx}	
\usepackage{amsmath}	

\title[A light echo in F01004-2237]{A light echo from the warm outflow in the ULIRG F01004-2237 following a major flare in its optical continuum emission}

\author[C. N. Tadhunter et al.]{
C. Tadhunter$^{1}$\thanks{E-mail: c.tadhunter@sheffield.ac.uk},
M. Patel$^{1,2}$, J. Mullaney$^{1}$
\\
$^{1}$ Department of Physics \& Astronomy, University of Sheffield, Sheffield S6 3TG, UK\\
$^{2}$ Department of Physics \& Astronomy, University of Leicester, University Road, Leicester LE1 7RH
\\
}

\pubyear{2021}

\hypersetup{draft}
\begin{document}
\label{firstpage}
\pagerange{\pageref{firstpage}--\pageref{lastpage}}
\maketitle

\begin{abstract}
Emission-line variability studies have the potential to provide key information about the structures of the near-nuclear outflow regions of AGN.
Here we present a VLT/Xshooter spectrum of the nucleus of the ULIRG F01004-2237 that was taken in August 2018,  about 8\,yr after a major flare in its integrated
optical emission. Compared with our WHT/ISIS spectrum from September 2015, the broad, red wings of the emission lines most closely associated with the flaring event, including HeII$\lambda$4686, NIII$\lambda\lambda$4640,4100 and HeI$\lambda$5876, have substantially declined in flux. In contrast, the broad, blue wings that dominate the [OIII], [NeIII], [NeV] and [OI] forbidden lines have increased in flux by a factor 1.4 -- 4.4 (depending on the line). Moreover, the [FeVII]$\lambda$6087 line is detected in the new spectrum for the first time. We interpret these results in terms of a light echo from the outflowing warm gas: the direct emission from the flaring event is continuing to fade, but due to light travel time effects we are only now observing the impact of the flare on the emission from the extended outflow region. Unless the outflow is confined to a small range of angles close to our line of sight, these observations imply that the outflow must be relatively compact ($r <  50$\,pc). In terms of the nature of the flare event, we speculate that the properties of the F01004-2237 flare may be the result of a tidal disruption event (TDE) occurring in an object with pre-existing AGN activity. 

\end{abstract}




\section{Introduction}

A major focus of recent studies of active galactic nuclei (AGN) has been on the outflows they drive, since such outflows may regulate star formation, and thereby affect the evolution of the host galaxies \citep[e.g.][]{veilleux05,king15}. However, the true impact of the outflows has proved hard to quantify, due to major uncertainties about their basic properties, such as spatial extents and densities \citep[see discussion in][]{rz13,harrison18}.

Hydrodynamic simulations suggest that AGN-driven outflows may be particularly important in the final stages of major galaxy mergers, as both the central galaxy bulges and super-massive black holes (SMBH) grow rapidly through the radial gas flows induced by the mergers \citep[e.g.][]{dimatteo05,johansson09}. Therefore, ultra-luminous infrared galaxies \citep[ULIRGs][]{sanders96}, which represent the peaks of major, gas-rich mergers, close to the time of coalescence of the SMBH from the two merging galaxies, are key objects for investigating the importance of AGN-driven outflows. Indeed, many studies have revealed outflows in ULIRGs across a range of phases of the ISM, including molecular, neutral and warm ionized  phases \citep[e.g.][]{rupke13,rz13,cicone14,gonz17,morganti15,veilleux17}. There is also evidence that the warm ionized outflows in ULIRGs with optically detected AGN are more extreme in their kinematic properties than those detected in non-ULIRG AGN \citep{rz13,arribas14}.

Moving beyond simply detecting the AGN-induced outflows, we have undertaken a programme to precisely quantify their physical properties for the first time using observations of a 90\% complete sample of 15 nearby ULIRGs with AGN nuclei detected at optical wavelengths \citep{rose18,spence18,tadhunter18,tadhunter19}. In contrast to some previous studies of warm outflows in AGN, our results show that the near-nuclear outflows in ULIRGs are relatively compact \citep{tadhunter18,tadhunter19} and have high densities \citep{rose18,spence18}. The resulting mass outflow rates and kinetic powers fall at the lower end or below the range of predictions from the hydrodynamic simulations \citep{rose18,spence18,tadhunter19}. 

Despite this progress, considerable uncertainties remain about the spatial structures of the warm outflow regions. Unfortunately, few ULIRGs have outflows that are large enough to be well-resolved in Hubble Space Telescope (HST) observations \citep{tadhunter18}. Therefore, it is important to develop alternatives to direct imaging in order to map the outflows. One possibility involves measuring the response of the outflow emission to rapid changes in the ionizing continuum -- a technique that has been successfully used, for example, to investigate the structure and radial extent of the broad line regions (BLR) of AGN \citep[reverberation analysis:][]{peterson93}. However, application of this technique to the outflows in the narrow line region (NLR) is challenging, and depends on there being sufficient warm gas present on small radial scales, 
as well as the occurrence a well-defined continuum flare in the illuminating AGN. Indeed, the detection of significant variability in the stronger forbidden lines (e.g. [OIII]) is rare in the general population of AGN, with few convincing cases reported in the literature (e.g. 3C390.3: Zheng et al. 1995; NGC5548: Peterson et al. 2013). Although major changes in the forbidden-line fluxes (incl. [OIII]) have been detected for a group of objects with unusually strong coronal emission lines (e.g. [FeVII] and [FeX]) in their SDSS spectra, this has been attributed to the effect of tidal disruption event (TDE) flares, rather than AGN \citep{komossa09,wang11,yang13}.

The subject of this paper -- the ULIRG F01004-2237 -- is a promising candidate for such studies due to its variability properties. Most notably, despite appearing as a type II AGN in spectra taken in 2000 and 2007, it displayed unusually strong and broad ($FWHM > 3000$\,km s$^{-1}$) HeII, HeI and NIII emission lines in a spectrum taken in 2015. Subsequently, examination of archival Catalina Sky Survey (CSS) light-curve data revealed that the source had undergone a luminous continuum flare in 2010. Based on the detection of this flare and the strong He emission lines, which are sometimes observed in TDEs, we identified F01004-2237 as a TDE \citep{tadhunter17}. However, viewed in this way, F01004-2237 is unusual in its properties. In particular, whereas in most TDE candidates the TDE-related features fade away and become invisible on a timescale of $\sim$1\,yr or less, in F01004-2237 the broad He features were still present in our 2015 spectrum, approximately 5\,yr after the original flare. Moreover, the light curve of F01004-2237 showed signs of late-time flattening, rather than the expected steep decline. Recently, \citet{trakhtenbrot19} have proposed an alternative explanation for the variability of F01004-2237 in terms of AGN activity. 


Here we present the analysis of a deep VLT/Xshooter spectrum of F01004-2237 taken in August 2018 -- about 8\,yr after the original flare -- that has a direct bearing on our understanding of both its AGN outflows, and the nature of its flaring activity. The paper is organised as follows. After reviewing previous studies of F01004-2237 in Section 2, we detail the observations and data reduction  in Section 3, and the main results on the variability of the emission lines are presented in Section 4. These results are then used to investigate the properties of the warm outflow in Section 5, which also includes discussion of the nature of the 2010 flaring event. Finally, the conclusions are presented in Section 6. 

For our assumed  cosmology with H$_{0}$ = 73 km s$^{-1}$, $\Omega_{\rm m}$ = 0.27 and
$\Omega_{\Lambda}$ = 0.73, 1\,arcsec corresponds to  2.03\,kpc
at the redshift of F01004-2237 ($z=0.11783\pm 0.00009$), as derived from fits to extended, narrow emission lines on either side of the nucleus by \citet{spence18}.

\begin{table*}
\centering
\begin{tabular}{|llllll|}
\hline\hline
Line ID &Component &Velocity &FWHM &Flux/$10^{-15}$ &(VLT2018)/ \\
Lab $\lambda$ (\AA) & &(km s$^{-1}$) &(km s$^{-1}$) &(erg cm$^{-2}$ s$^{-1}$) 
&(WHT2015) \\
\hline
OIII(3133) &i &-119$\pm$18 &562$\pm$59 &0.28$\pm$0.04 & \\
	   &b &-447$\pm$30 &2060$\pm$85 &1.36$\pm$0.55 & \\
\hline
[NeV](3426)  &i  &-219$\pm$5 &794$\pm$26 &1.75$\pm$0.13 &  \\
             &b  &-815$\pm$50 &1610$\pm$50 &1.97$\pm$0.14 & \\
	     &i$+$b & & &3.72$\pm$0.19 &0.71$\pm$0.05 \\
\hline
[NeIII](3869) &n &-25.3$\pm$1.3 &108$\pm$2 &0.33$\pm$0.01 & \\
	      &i &-264$\pm$10 &879$\pm$41 &1.53$\pm$0.24 & \\
	      &b &-814$\pm$84 &1690$\pm$70 &1.29$\pm$0.27 & \\
	      &i$+$b & & &3.82$\pm$0.36 &1.42$\pm$0.17 \\
\hline
NIII(4640)  &b &-78$\pm$11 &1540$\pm$30 &1.03$\pm$0.02 & \\
            &i &-14$\pm$17 &661$\pm$43 &0.20$\pm$0.02 & \\
HeII(4686) &b1 &-303$\pm$21 &1620$\pm$50 &1.27$\pm$0.04 & \\
            &n &-38$\pm$7 &382$\pm$27 &0.15$\pm$0.02 & \\
	    &b2 &139$\pm$65 &2240$\pm$160 &0.55$\pm$0.04 & \\
	    &Total & & &3.20$\pm$0.07 &0.50$\pm$0.02 \\
\hline
H$\beta$(4861) &n &-28.5$\pm$0.3 &95$\pm$1 &0.87$\pm$0.01 &\\
	       &b1 &-261$\pm$16 &1100$\pm$20 &2.46$\pm$0.08 & \\
	       &b2 &-1300$\pm$60 &1100$\pm$90 &0.60$\pm$0.08 & \\
	       &b1$+$b2 & & &3.06$\pm$0.11 &0.99$\pm$0.05  \\
\hline
[OIII](5007)   &n &-25.1$\pm$0.2 &78$\pm$1 &3.35$\pm$0.03 & \\
	       &i &-349$\pm$8 &818$\pm$22 &5.85$\pm$0.44 & \\
	       &b &-1170$\pm$40 &1480$\pm$60 &7.32$\pm$0.47 & \\
	       &i$+$b & & &13.2$\pm$0.6 &1.50$\pm$0.10 \\
\hline
HeI(5876)      &n &-110$\pm$1 &99$\pm$3 &0.127$\pm$0.004 & \\
	       &i &-192$\pm$12 &877$\pm$26 &0.56$\pm$0.04 & \\
	       &b &-1080$\pm$110 &1430$\pm$160 &0.25$\pm$0.04 & \\
	       &i$+$b & & &0.81$\pm$0.06 &0.44$\pm$0.04 \\
\hline
[OI](6300)     &n &-31$\pm$1 &152$\pm$2 &0.222$\pm$0.03 & \\
	       &i &-273$\pm$1 &709$\pm$33 &0.18$\pm$0.01 & \\
	       &b &-596$\pm$1 &1700$\pm$40 &0.66$\pm$0.03 & \\
	       &i$+$b & & &0.84$\pm$0.03 &1.83$\pm$0.24  \\
\hline
[FeVII](6087)  &i &-229$\pm$17 &794$\pm$15 &0.19$\pm$0.01 & \\
	       &b &-825$\pm$32 &1607$\pm$28 &0.24$\pm$0.02 & \\
\hline
HeI(10833)     &n &-27$\pm$1 &194$\pm$3 &1.27$\pm$0.27 & \\
	       &i &-205$\pm$6 &946$\pm$24 &4.35$\pm$0.26 & \\
	       &b &-366$\pm$14 &2006$\pm$68 &4.73$\pm$0.24 & \\
\hline
Pa$\alpha$(18756)     &n1 &-9$\pm$1 &142$\pm$1 &3.89$\pm$0.03 & \\
	             &n2 &-75.0$\pm$5 &444$\pm$14 &1.15$\pm$0.04 & \\
	             &b &-555$\pm$16 &1976$\pm$40 &2.82$\pm$0.06 & \\
\hline\hline
\end{tabular}
\caption{Fluxes measured for the various kinematic components fitted to some of  the brighter emission lines 
detected in the 2018 Xshooter spectrum (column 5). In each case, a combination of narrow (n), intermediate (i), and broad (b)
kinematic components were fitted to the lines;  the rest-frame velocity shifts and instrumentally-corrected rest-frame FWHM are 
reported in columns 3 and 4 respectively. The final column (column 6) compares
the total fluxes measured for the intermediate and broad components from the 2018
Xshooter spectrum with those measured from the 2015 WHT/ISIS spectrum. Note that
in the case of the NIII$\lambda$4640+HeII$\lambda$4686 blend, the individual
kinematic components of the two contributing species are difficult to separate, so
when comparing with the 2015 WHT/ISIS spectrum we use the total flux for the 
blend as a whole. }
\label{table:fluxes}
\end{table*}

\vglue -1.0cm\noindent
\begin{figure}
\includegraphics[width=10.0cm]{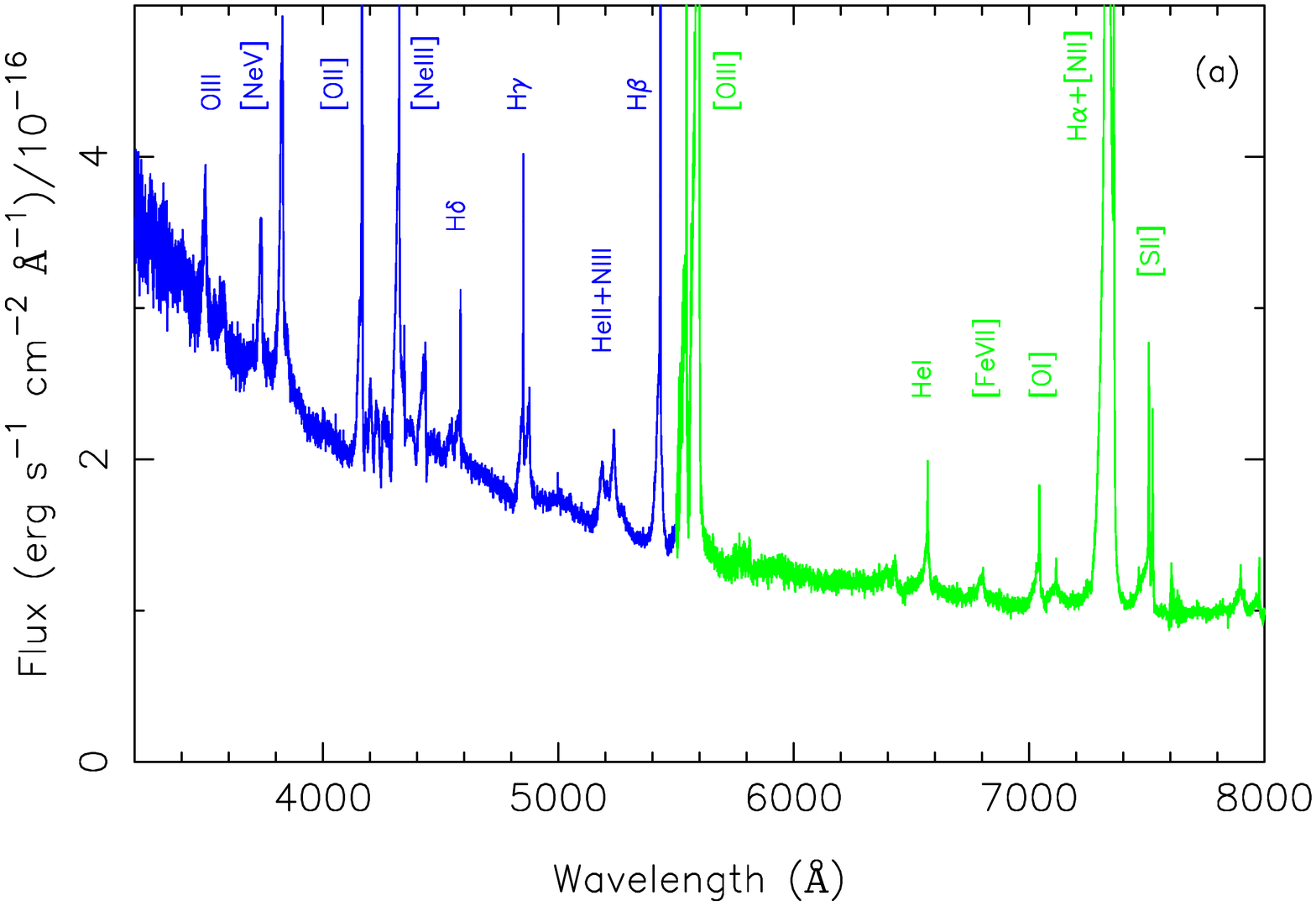}
\includegraphics[width=10.0cm]{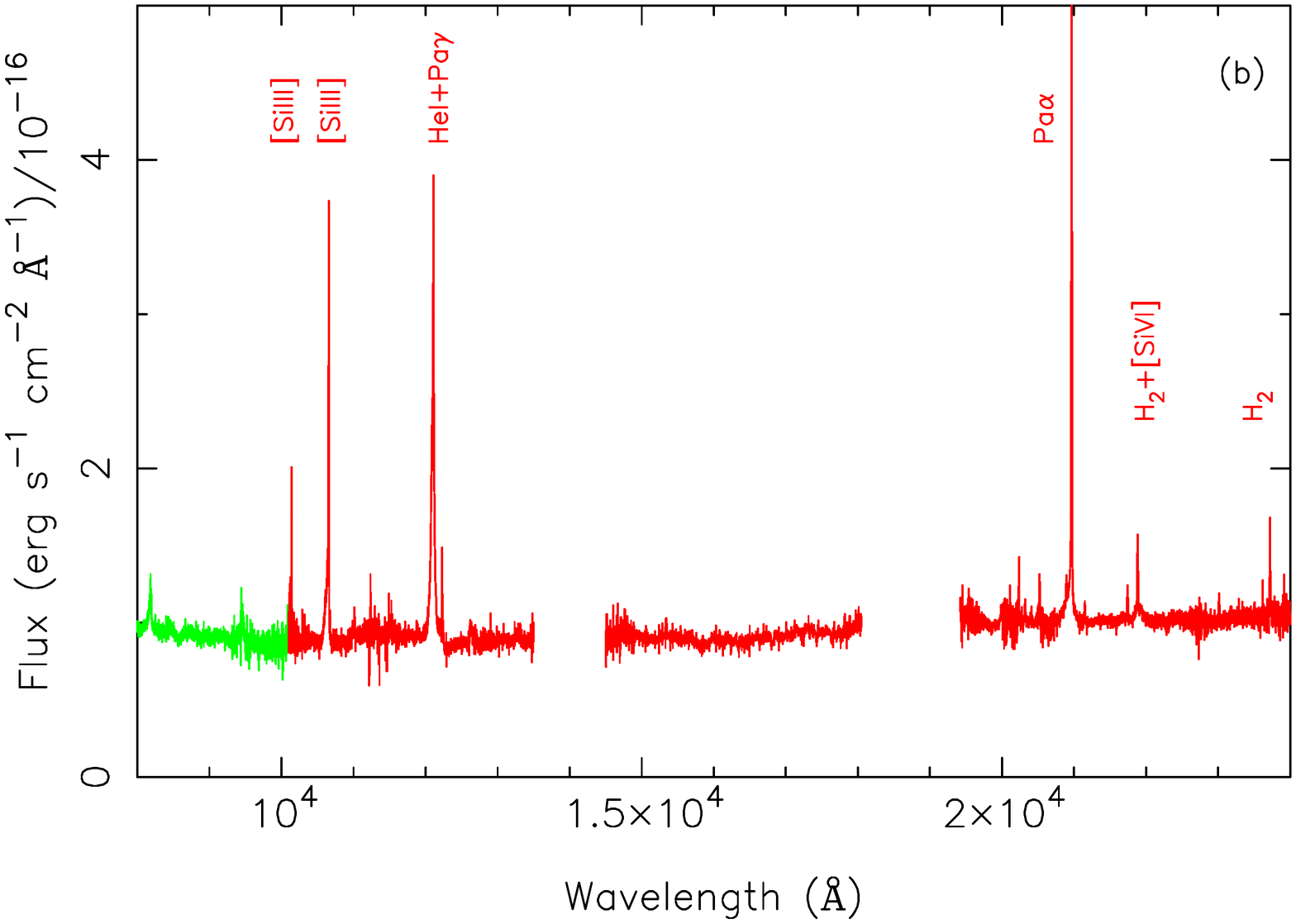}
\caption{The full 2018 X-shooter spectrum of F01004-2237 from UV to near-IR wavelengths: (a) UV to optical wavelengths; (b) near-IR wavelengths. The UVB, VIS and IR spectra
are shown in blue, green and red respectively. Note the steep rise in the
continuum at the blue/UV at wavelengths $\lambda < 0.6$\,$\mu$m. The gaps the near-IR spectrum at $\sim$14,000 and  $\sim$19,000\,\AA\,
represent regions that are strongly affected by telluric absorption. The wavelengths are observed-frame rather than rest-frame.}
\label{fig:full_xshooter}
\end{figure}

\section{Previous work on F01004-2237}

F01004-2237 is classified as a ULIRG based on its far-IR luminosity ($L_{IR} = 1.9\times10^{12}$\,L$_{\odot}$), and is part of the 1Jy sample of ULIRGs described in \citet{kim98}. Emission-line fluxes and line ratios for the stronger emission lines in its optical nuclear spectrum are presented in  \citet{veilleux99}, \citet{farrah05} and \citet{rz13}.  Based on BPT diagnostic diagram analysis of line ratios derived from integrated line fluxes, there was initially some uncertainty about the classification of its nuclear spectrum, with  \citet{veilleux99} adopting an HII classification and \citet{yuan10} a Seyfert 2 classification. The reasons for this apparent ambiguity became clear when the line ratios of individual kinematic components were analysed by \citet{rz13}: whereas the line ratios of the broad kinematic component are consistent with AGN photoionisation and a Seyfert 2 classification, those of the narrowest kinematic component suggest an HII classification, and therefore the presence of kinematically quiescent gas photoionized by hot stars. 

Further evidence for young stellar populations in this source is provided by the detection of broad Wolf-Rayet features in the optical spectrum by \citet{armus88} and \citet{lipari03}, and near-nuclear star clusters in the HST images presented by \citet{surace98}; young stellar ages ($t<$10\,Myr) are suggested by the presence of the Wolf-Rayet features, as well as by the colours of the star clusters and the nucleus itself, which is barely resolved in B-band HST images \citep{surace98}.

F01004-2237 was one of the 15 ULIRGs with optical AGN observed as part the study of warm outflows in ULIRGs described in the introduction 
\citep{rz13,rose18,spence18}. Comparison of spectra taken in 2007 and 2015 as part of this study provided the first evidence for emission-line variability. Subsequently, an optical  continuum flare -- peaking in 2010 -- was detected in the CSS light curve for the source \citep{tadhunter17}, followed by the detection of a light echo from the circum-nuclear dust at mid-IR wavelengths \citep{dou17}.  As discussed in detail in \citet{tadhunter17}, if the flare represents a TDE, the detection of one such flare in a sample of only 15 ULIRGs over a timescale of $\sim$8 years, implies that the rate of TDEs is much higher in the major mergers represented by ULIRGs than in the general population of galaxies.

Regardless of whether the 2010 flare in F01004-2237 represents a TDE or an unusual form of AGN activity \citep{trakhtenbrot19}, its detection provides evidence that we currently have an unobscured view of the nuclear regions close to the SMBH, {which is estimated to have a mass of $2.54\times10^7$\,M$_{\odot}$, based on the stellar velocity dispersion of the system \citep{dasyra06}}. In this context it is interesting that, prior to 2015, F01004-2237 was classified as type II AGN, with no reports of the strong, broad ($FWHM > 2000$\,km s$^{-1}$) permitted emission lines that would be expected of a classical type I AGN. Moreover, although weak X-ray emission has been detected in Chandra observations of F01004-2237 \citep[$L_{\rm 2-10\,keV} = 1.3\times10^{42}$\,erg s$^{-1}$;][]{teng10}, it has been argued by \citet{nardini11} that the properties of this X-ray emission are consistent with a starburst rather than an AGN origin. Therefore, given the clear evidence for type II AGN activity from the line ratios of the broad kinematic components of the optical emission lines \citep{rz13}, it is possible that prior to the flare in 2010 the AGN had recently entered a low activity state. Note, however, that it is difficult to entirely rule out the presence of emission from a type I AGN  at some level in spectra taken before 2015, because  emission lines from the BLR could be masked by the  strong, broad emission-line wings associated with the warm outflow detected in the forbidden lines (see section 4.2 below), and the direct AGN continuum could be difficult to distinguish from the light of the young stellar population. This is especially the case given that spectra taken before 2015 had either a lower S/N \citep[e.g.][]{rz13} or lower spectral resolution \citep[HST spectrum in][]{farrah05} than
those presented in this paper and in \citet{tadhunter17}.

\section{Observations and reductions}

The data for this project were taken using the Xshooter instrument \citep{vernet11} on the ESO VLT. The conditions were photometric for the observations, which were made in two separate observing blocks on the nights of the 12th and 13th of August 2018 with the slit aligned along PA90. Slit widths of 1.3, 1.2 and 1.2 arcsec were used for the UVB, VIS and IR arms of Xshooter, resulting in instrumental FWHM of 1.0, 1.1 and 2.4\AA, respectively. The combined total on-source integration times for the observations were 5040, 5040 and 5280\,s for observations in the UVB, VIS and IR arms of Xshooter respectively. 

The data for the two nights were reduced using the standard Xshooter pipeline running under EsoRex\footnote{https://www.eso.org/sci/software/cpl/esorex.html}. 1D spectra were extracted from the rectified, wavelength and flux calibrated long-slit spectra using an extraction aperture of length 1.8 arcsec in the spatial direction and centred on the nucleus of F01004-2237. Then the spectra were corrected for telluric absorption features using the Molecfit package. Finally, the 1D nuclear spectra from the two nights were averaged to produce the spectra that we will use for analysis in the remainder of this paper. The full 2018 Xshooter spectrum from UV to near-IR wavelengths is shown in Figure \ref{fig:full_xshooter}. This demonstrates an excellent match in flux levels  between spectra taken separately using the UVB, VIS and IR arms of Xshooter. 

We used the target acquisition images to estimate the effective along-the-slit seeing and slit losses for the data taken on the two nights. In both acquisition images -- taken using a Sloan g' filter -- we identified two stars positioned on either side of F01004-2237 that could be used for this purpose. In each case, we extracted spatial slices with the same width (1.2 arcsec) as the spectroscopic slit and centred on the stars. We then fitted Gaussians  to the resulting spatial profiles using the STARLINK DIPSO package, in order to determine the full widths at half maximum (FWHM) and total fluxes contained within the profiles.  Averaging the results for the two stars and both nights we obtained $FWHM = 0.90\pm0.04$, which we regard as a good estimate of the effective 1D (i.e. along-the-slit) seeing for the observations. This process was also repeated using spatial slices extracted with a much larger width in the direction perpendicular to the direction of the slices (5 arcsec), in order to determine the slit losses by comparing the fluxes of the stars measured from the narrow (1.2 arcsec) and wide (5 arcsec) spatial slices. In this way, we estimated that the slit losses for the 2018 VLT/Xshooter observations are in the range 16 -- 25\%. This is similar to the slit losses for the 2016 WHT/ISIS observations (26$\pm$3\%) and the 2000 HST/STIS observations (10 -- 20\%) of F01004-2237, as estimated in \citet{tadhunter17}.

\begin{figure}
\includegraphics[width=9.5cm]{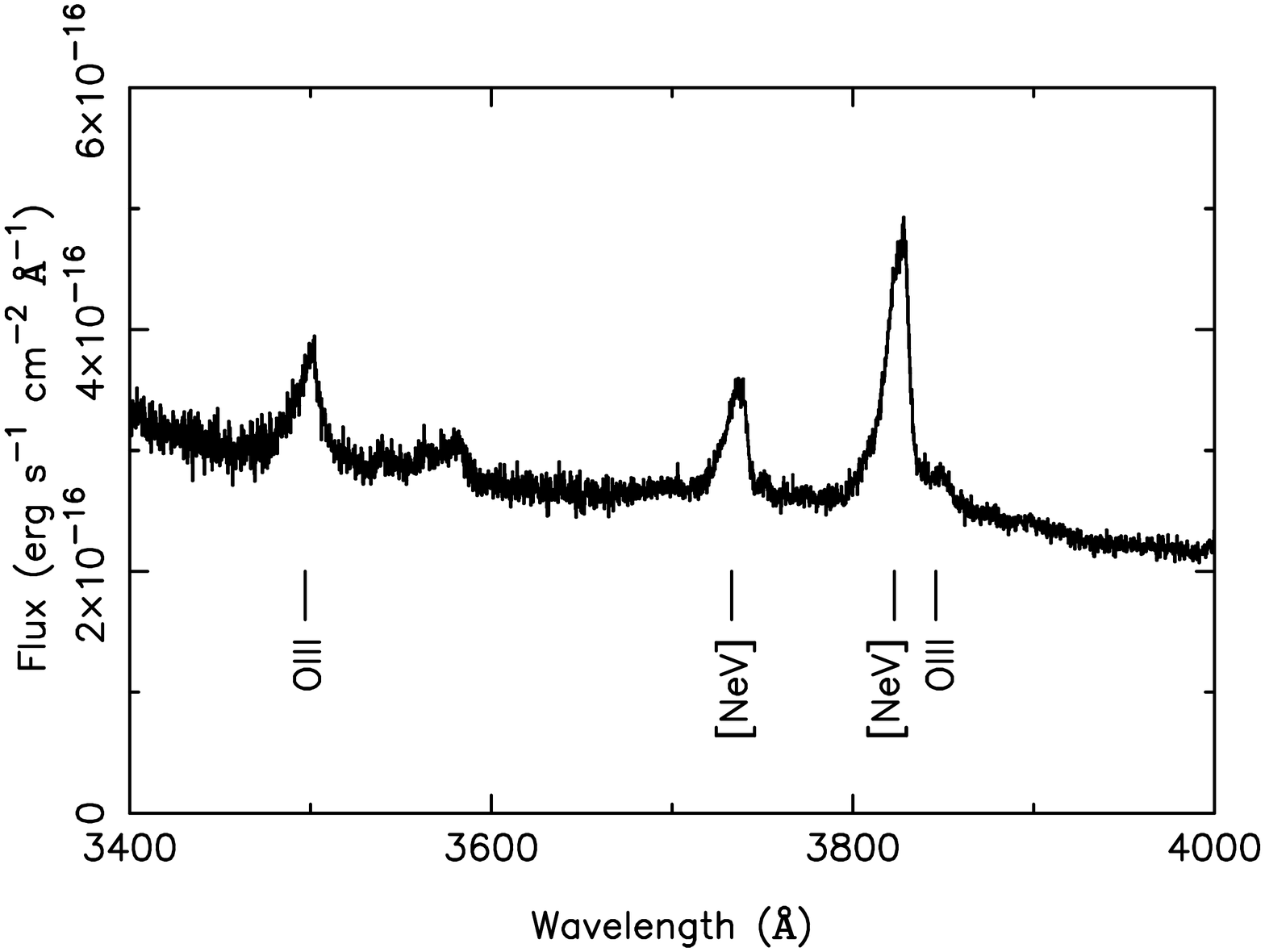}
\caption{UV spectrum of F01004-2237 showing the clear detection of the OIII$\lambda$3133,3444 Bowen lines as well as the
[NeV]$\lambda\lambda$3346,3426 doublet.}
\label{fig:bowen}
\end{figure}

\section{Results}

\subsection{Outflow-related features}

\begin{figure}
\includegraphics[width=9.5cm]{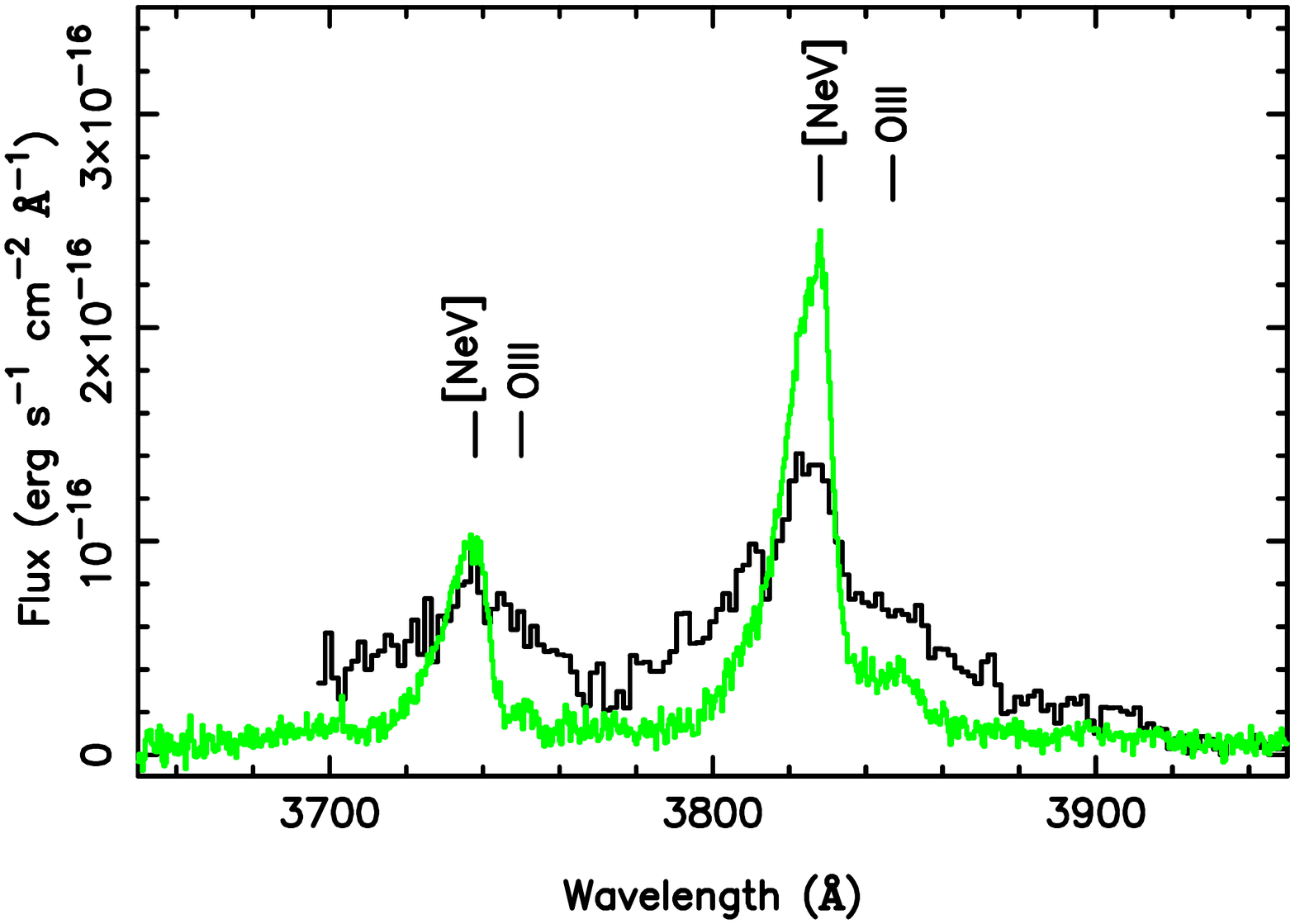}
\caption{Comparison between the 2015 (black) and 2018 (green) spectra for a wavelength region containing the [NeV]$\lambda\lambda$3346,3426 doublet lines. Note that the broad wings to the features visible in the 2015 spectrum are likely to be due to OIII$\lambda\lambda\lambda$3341,3429,3444 Bowen lines, and these wings have substantially declined between 2015 and 2018. On the other hand, the cores of the features -- dominated by broad and intermediate, blueshifted components of the [NeV]$\lambda\lambda$3346,3426 lines -- have substantially increased over the same period. For each epoch, a simple power-law continuum has been subtracted, but no scaling of the fluxes has been applied.}
\label{fig:nev}
\end{figure}

The UV-optical part of the full 2018 VLT/Xshooter spectrum shows strong similarities with the 2015 WHT/ISIS spectrum presented in \citet{tadhunter17}, including the detection of broad, blue shifted forbidden lines of [NeV], [OII], [NeIII], [OIII], [OI], [NII] and [SII], along with broad HI, HeI and HeII permitted lines. However, the more extended spectral range of the Xshooter observations to the UV has also allowed us to clearly detect  the OIII$\lambda$3133 Bowen resonance-fluorescence line for the first time. Given the presence of the latter feature, we identify the bump in the red wing of the [NeV]$\lambda$3425 line as the OIII$\lambda$3444 Bowen resonance-fluorescence line (see Figure \ref{fig:bowen}); the presence of this line -- which is likely to have been stronger closer to the time of the continuum flare -- helps to explain the apparently peculiar profile of the [NeV]$\lambda$3426 feature in our 2015 WHT/ISIS spectrum (see Figure \ref{fig:nev}). We further note that the nuclear continuum SED rises steeply to the UV (Figure \ref{fig:full_xshooter}). However, it is not currently clear whether this is due to a TDE, an AGN, or the presence of a young stellar population. Certainly, the detection of  Balmer break and the higher order Balmer lines in absorption, albeit relatively weak, shows that young stars must contribute to the near-UV continuum at some level.

\begin{figure}
\includegraphics[width=9.5cm]{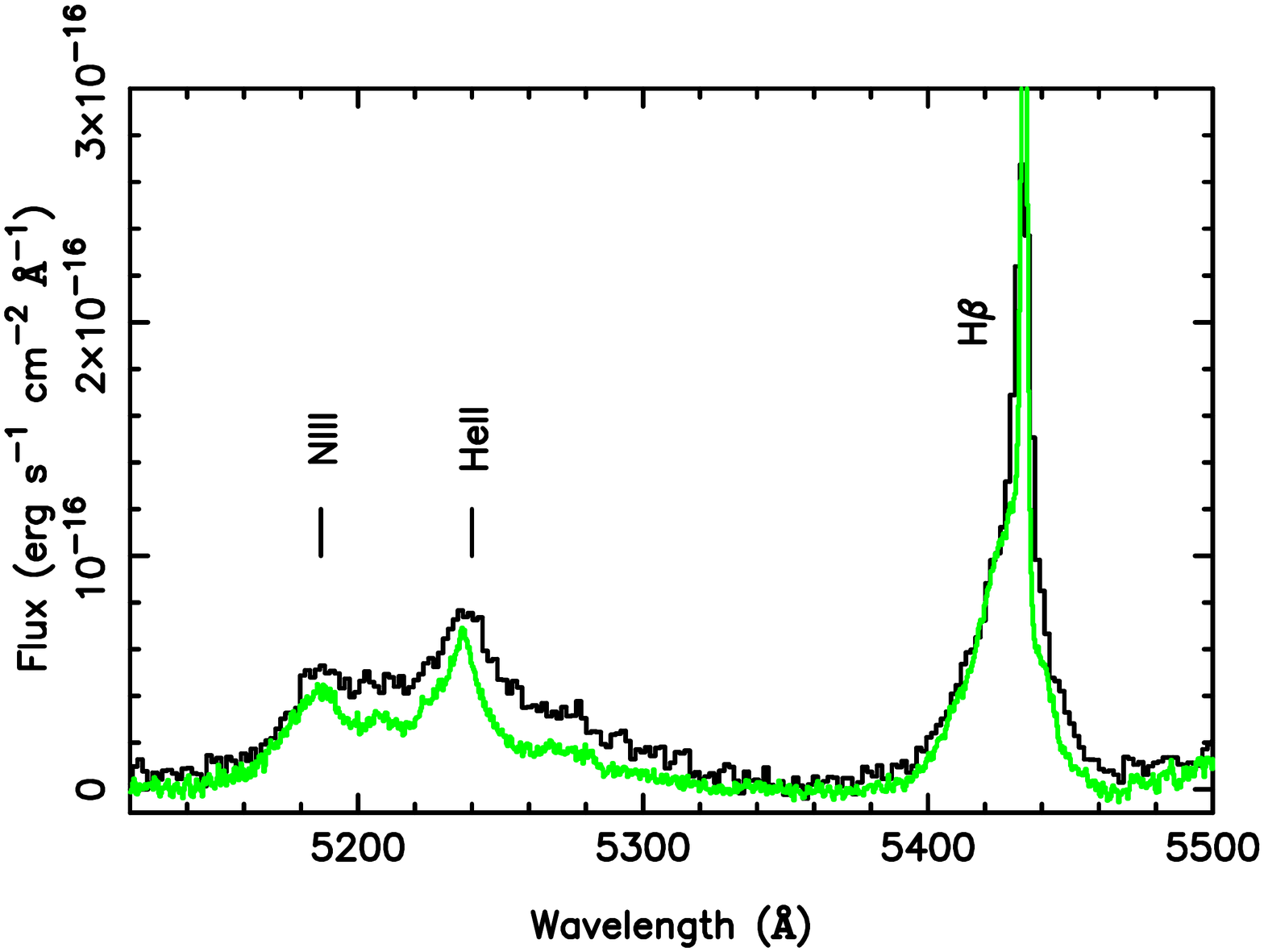}
\includegraphics[width=9.5cm]{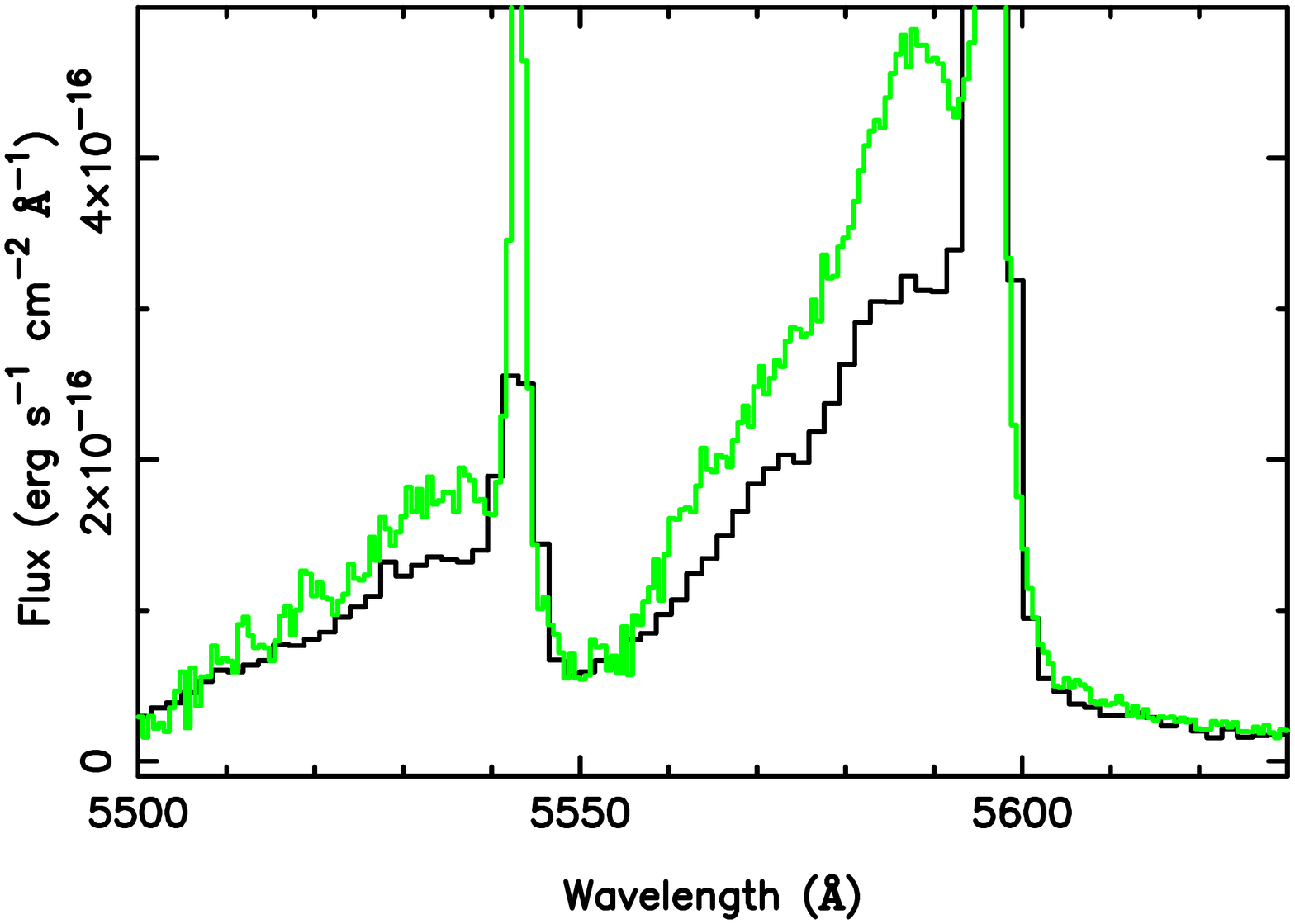}
\caption{Comparison between the 2015 (black) and 2018 (green) spectra for blue/green wavelength regions containing the HeII+NIII and H$\beta$ permitted lines (top), and  [OIII]$\lambda\lambda$4959,5007 forbidden lines (bottom). In the case of each epoch, a simple power-law continuum has been subtracted, but no scaling of the fluxes has been applied.}
\label{fig:blue}
\end{figure}

\begin{figure}
\includegraphics[width=9.5cm]{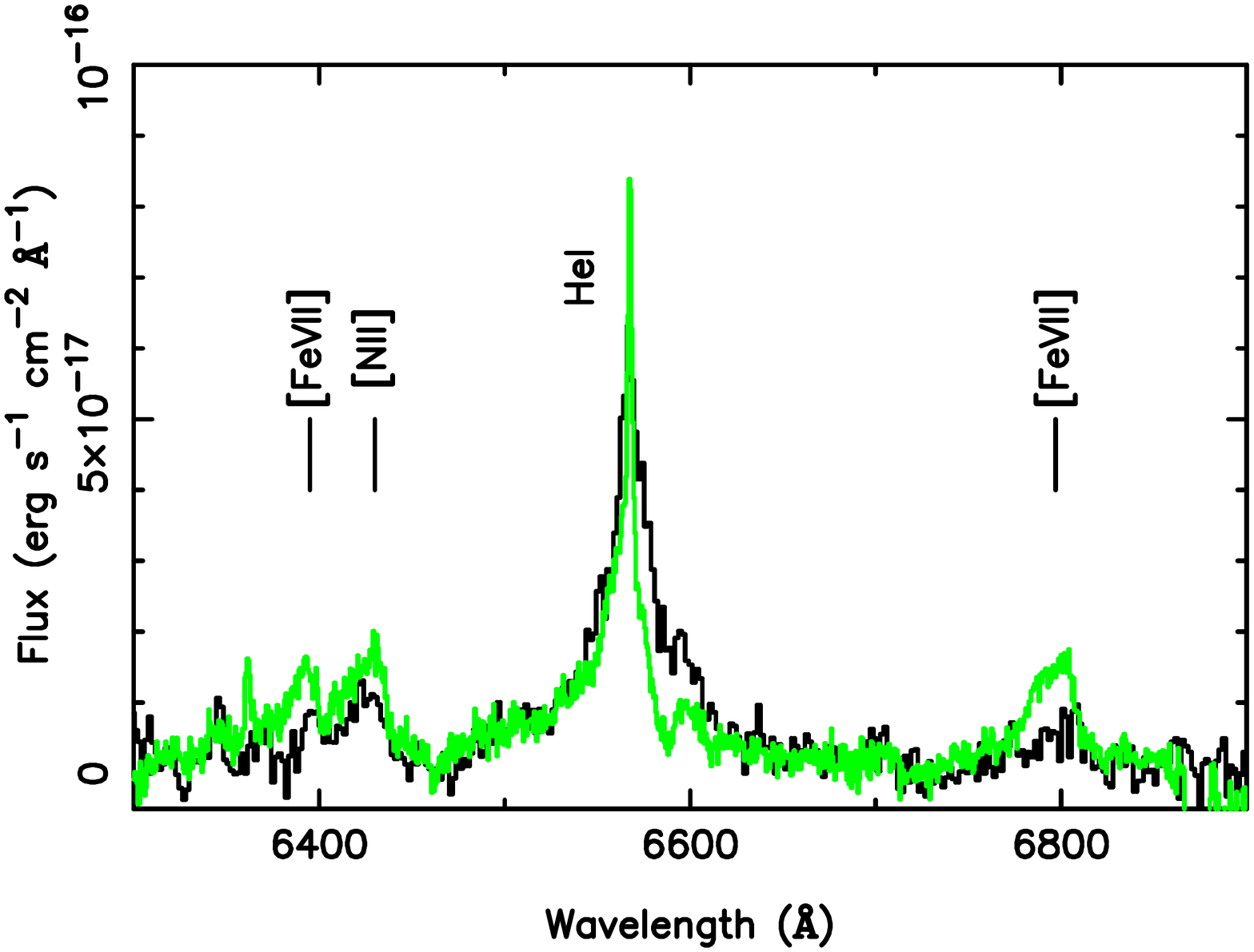}
\includegraphics[width=9.5cm]{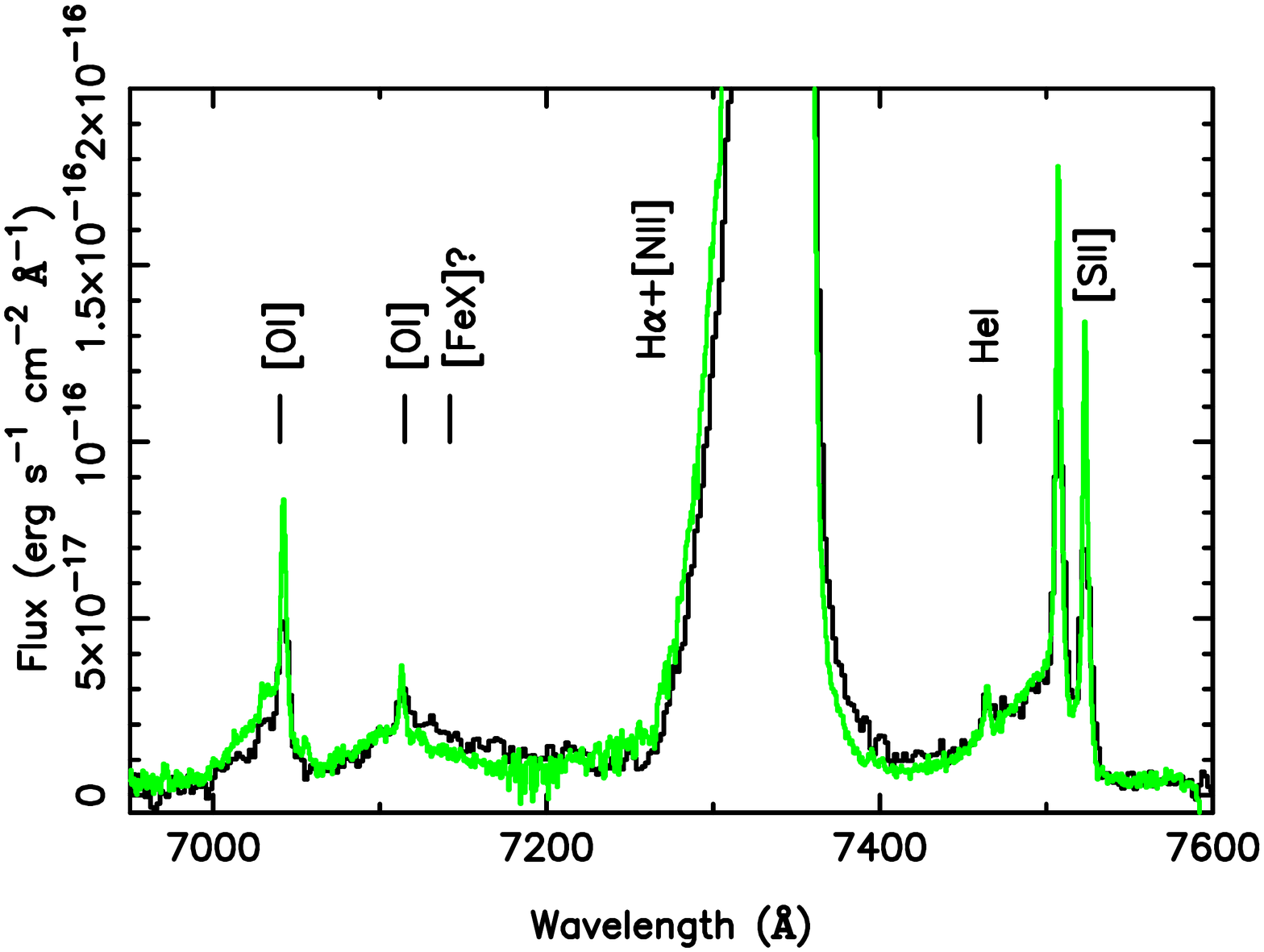}
\caption{Comparison between the 2015 (black) and 2018 (green) spectra for red wavelength regions containing the  HeI$\lambda$5876 and [FeVII]$\lambda$6087 lines (top), and blended  [OI]$\lambda\lambda$6300,6363, [NII]$\lambda$6548, H$\alpha$, [NII]$\lambda$6584, [SII]$\lambda\lambda$6717,6731  lines (bottom). Note the decline in the red wing of the HeI$\lambda$5876 feature and simultaneous increase in the strength of the [FeVII]$\lambda$6087 feature between 2015 and 2018. In the case of each epoch, a simple power-law continuum has been subtracted, but no scaling of the fluxes has been applied.}
\label{fig:red}
\end{figure}

Our near-IR Xshooter spectrum -- the first published for this object -- shows the presence of broad [SIII], HeI, HI emission features, consistent in terms of their kinematics with trends observed in the features detected at UV-optical wavelengths.

Despite the superficial similarities, the 2018 VLT/Xshooter spectrum shows major differences from the 2015 WHT/ISIS spectrum. These become apparent when we compare the line fluxes (Table \ref{table:fluxes}) and detailed emission line profiles (Figures \ref{fig:blue}, \ref{fig:red} and \ref{fig:nev}). The line fluxes were determined by fitting multiple Gaussian profiles to emission lines using the routines in DIPSO, which is part of the STARLINK software package.  During the fitting, a simple power-law shape was assumed for the underlying continuum; given the lack of structure and of detectable absorption lines free from emission line contamination, it was not deemed appropriate to perform spectral synthesis modelling of the
continuum.  In the case of blends involving emission-line doublets arising from the same upper energy level, we fixed the flux ratios and wavelength separations of emission line doublets to their atomic physics values corrected for redshift, and required the widths of particular kinematic components to be the same for the two doublet lines. All the emission lines require a combination of intermediate ($500 < FWHM < 1000$\,km s$^{-1}$) and broad ($FWHM > 1000$\,km s$^{-1}$) kinematic components, and most also require a narrow component ($FWHM <$500\,km s$^{-1}$) to be adequately fit; as already clear from previous studies of this object, the broad, blue wings of the [OIII] emission lines are particularly striking, extending to blue shifted velocities $\Delta V = -2500$\,km s$^{-1}$. However, all the emission lines show broad wings.

In Table \ref{table:fluxes} we present the fluxes of the broad and intermediate emission line components measured from our 2018 VLT/Xshooter spectrum, and compare them with those measured from the 2015 WHT/ISIS spectrum presented in Tadhunter et al. (2017). Note that we do not compare the fluxes of the narrow kinematic components, because the narrow lines are clearly spatially extended in the long-slit spectra, and their fluxes are expected to be different for the two epochs, due to the different slit PAs used for the observations. On the other hand, analysis of the spatial profiles along the slit for the broad/intermediate emission line features shows that they are not spatially resolved in our spectra, in the sense that the their spatial FWHM are similar to those measured for stars using simulated slits in the acquisition images. Given that the estimated slit losses (see Section 3) are similar for the 2015 and 2018 data, it is possible to directly compare the broad/intermediate fluxes between the two epochs. 

We find that the broad/intermediate components to the [NeIII], [OIII], [OI] and [SII] forbidden lines -- which represent the warm outflow -- have increased in flux by a factor of $\sim$1.4 -- 1.8 between 2015 and 2018. At the same time, the broad/intermediate components of the features most closely associated with the flaring event, including NIII$\lambda$4640, HeII$\lambda$4686, H$\beta$ and HeI$\lambda$5876 have fluxes measured from the 2018 spectrum that are similar to, or up to a factor $\sim$2 lower than, those measured in 2015 (see Section 4.2 below). Given that the continuum-subtracted spatial profiles of the wings of the forbidden and permitted lines are similar\footnote{Using Gaussian fits to the continuum subtracted spatial profiles of the emission line wings
we measure $FWHM = 0.89\pm0.03$\,arcec and $FWHM = 0.88\pm0.01$\,arcsec for the spatial profiles of the red wing ($+300 < \Delta V < 1400$\,km s$^{-1}$) of H$\beta$ and the blue wing ($-2000 < \Delta V < -300$\,km s$^{-1}$) of [OIII]$\lambda$5007 respectively; note that these spatial FWHM are similar to the effective seeing FWHM  of the observations (see Section 3).}, this latter result rules out the idea that the increase in the fluxes of the broad/intermediate wings of the forbidden is due to a failure to properly account for slit losses between the two epochs. It is also notable that in \citet{tadhunter17} we found no evidence for a significant change in the broad forbidden lines fluxes between the 2000 HST/STIS spectrum and the 2015 WHT/ISIS spectrum (discounting the [NeV]$\lambda$3426 feature which was likely contaminated by a Bowen resonance-fluorescence feature in 2015).

The trend of increasing flux in the broad/intermediate forbidden lines, and simultaneous decreasing flux  in the broad/intermediate component of some of the permitted lines, is also clear from the direct line profile comparisons shown in Figures \ref{fig:blue} and \ref{fig:red}. Note that we have not directly compared the fluxes and profiles of the strong [NII]$\lambda\lambda$6584,5648 and H$\alpha$ lines, because the extreme widths of the lines and overall kinematic complexity of the lines make it hard to confidently identify and separate the contributions of the different lines to the heavily blended H$\alpha$+[NII] feature.

The [NeV]$\lambda\lambda$3346,3426 lines require particular discussion. Although the  total flux contained in the broad/intermediate components of these features has apparently decreased since 2015, we note that the 2015 fluxes contain substantial contributions from the Bowen OIII$\lambda\lambda\lambda$3341,3429,3444 lines, which was not accounted for in the fits to the lower resolution WHT/ISIS spectrum. In contrast, the OIII$\lambda$3444 line was explicitly modelled in the fits to the [NeV]$\lambda$3426 line in the higher-resolution Xshooter spectrum, so the Xshooter [NeV]$\lambda$3426 flux has been corrected for the flux contribution of this line. The true situation is revealed by the profile comparison shown in Figure \ref{fig:nev}: while the broad red wing -- presumably contributed by the Bowen OIII$\lambda$3444 line -- declined in flux between 2015 and 2018, the strong core of the feature -- entirely contributed by the intermediate and broad blueshifted kinematic components of [NeV]$\lambda$3426  -- has substantially increased. Indeed the total flux measured in the broad/intermediate components of the 2018 Xshooter spectrum is a factor 4.4$\pm$0.7 higher than the total flux measured for this line from the 2000 HST/STIS spectrum\footnote{Note that we have not explicitly modelled the contribution of the Bowen OIII$\lambda$3429 line to the [NeV]$\lambda$3426 feature detected in the 2018 spectrum. However, the transition associated with this line shares the same upper level (2p3d\,\,$^3P^0_2$)  as that of the OIII$\lambda$3444 line, so the flux ratio of the two lines is set by the ratio of their transition probabilities. Using the transition probability ratio and the measured flux of the OIII$\lambda$3444 line, we estimate that the OIII$\lambda$3429 Bowen line contributes less than 5\% of the [NeV]$\lambda$3426 flux.}. Therefore, the [NeV]$\lambda$3426 line follows the same trend of increasing flux as those shown by the broad/intermediate blue shifted components of the other forbidden lines noted above, although in this case the flux increase is larger. 

It is also notable that the red wing of what we now identify as OIII$\lambda$3444 in the 2015 spectrum extends to redshifted velocities $\Delta v > 5000$\,km s$^{-1}$. This is similar to the maximum redshifted velocity measured for the HeII$\lambda$4686 line at the same epoch. Therefore, it is likely that these OIII and HeII lines originate in the same regions. 

Consistent with the [NeV]$\lambda\lambda$3346,3426 features, the broad [FeVII]$\lambda\lambda$5721,6087 lines -- which are emitted by ions with a similar ionisation potential to NeV -- have also increased markedly in flux. Indeed, these [FeVII] lines are clearly detected for the first time in our 2018 VLT/Xshooter spectrum (see Figure \ref{fig:red})\footnote{The [FeVII]$\lambda$5721 line might be weakly detected in the 2015 spectrum (see Figure \ref{fig:red}), but its identification at that epoch is uncertain, due to the lower resolution of the spectrum, the relatively low S/N in the continuum around the feature, and blending with the [NII]$\lambda$5755 line.}.

Finally, we note that the broad feature redward of the [OI]$\lambda$6363 line -- which we tentatively identified with [FeX]$\lambda$6374 in \citet{tadhunter17} -- appears to have significantly declined in flux between the two epochs. If the identification of the [FeX]$\lambda$6374 feature is correct, then this is the only broad forbidden line feature to have decreased in flux between 2015 and 2018. Interestingly, the apparent degrease in [FeX]$\lambda$6374 and simultaneous increase in [FeVII]$\lambda\lambda$5721,6087 is consistent with the trend noted by \citet{yang13} for two of their TDE candidate objects with unusually strong coronal emission lines.  However, some uncertainties remain about the detection of the [FeX]$\lambda$6374 feature in F01004-2237  because it is weak relative to the underlying continuum and the broad wings of the nearby H$\alpha$+[NII] blend.

Overall, the Xshooter observations provide strong evidence for an increase between 2015 and 2018 in the fluxes of the broad/intermediate components of forbidden lines covering a wide range of ionisation.

\subsection{Flare-related features}

In contrast to the forbidden lines, the broad/intermediate components of permitted lines that we have identified as being most closely associated with the flare event in 2010 have either substantially declined or remained constant in flux between 2015 and 2018. 

The flux decline amounts to factor of two or more in the NIII+HeII blend around 4600 -- 4750\,\r{A}, in the HeI$\lambda$5876 line and in the OIII$\lambda\lambda\lambda$3341,3429,3444 lines that are blended with [NeV]$\lambda\lambda$3346,3426. We further note that the broad wings to the feature at $\sim$4100\AA\, that was tentatively identified as a blend of H$\delta$ and NIII lines in \citet{tadhunter17},  are not detected in the new VLT/Xshooter spectrum.

Since we have now identified the OIII$\lambda\lambda$3133,3444 Bowen  lines in our VLT/Xshooter spectrum, we can also identify the NIII$\lambda\lambda$4100,4640 lines as being produced by the same resonance-fluorescence mechanism, thus supporting the interpretation of Trahktenbrot et al. (2019). From the evidence of the OIII$\lambda\lambda$3133,3444
and NIII$\lambda\lambda$4100,4640 features measured in our Xshooter spectrum, it is likely that all of these Bowen lines declined in flux between 2015 and 2018.

\begin{figure}
\includegraphics[width=9.5cm]{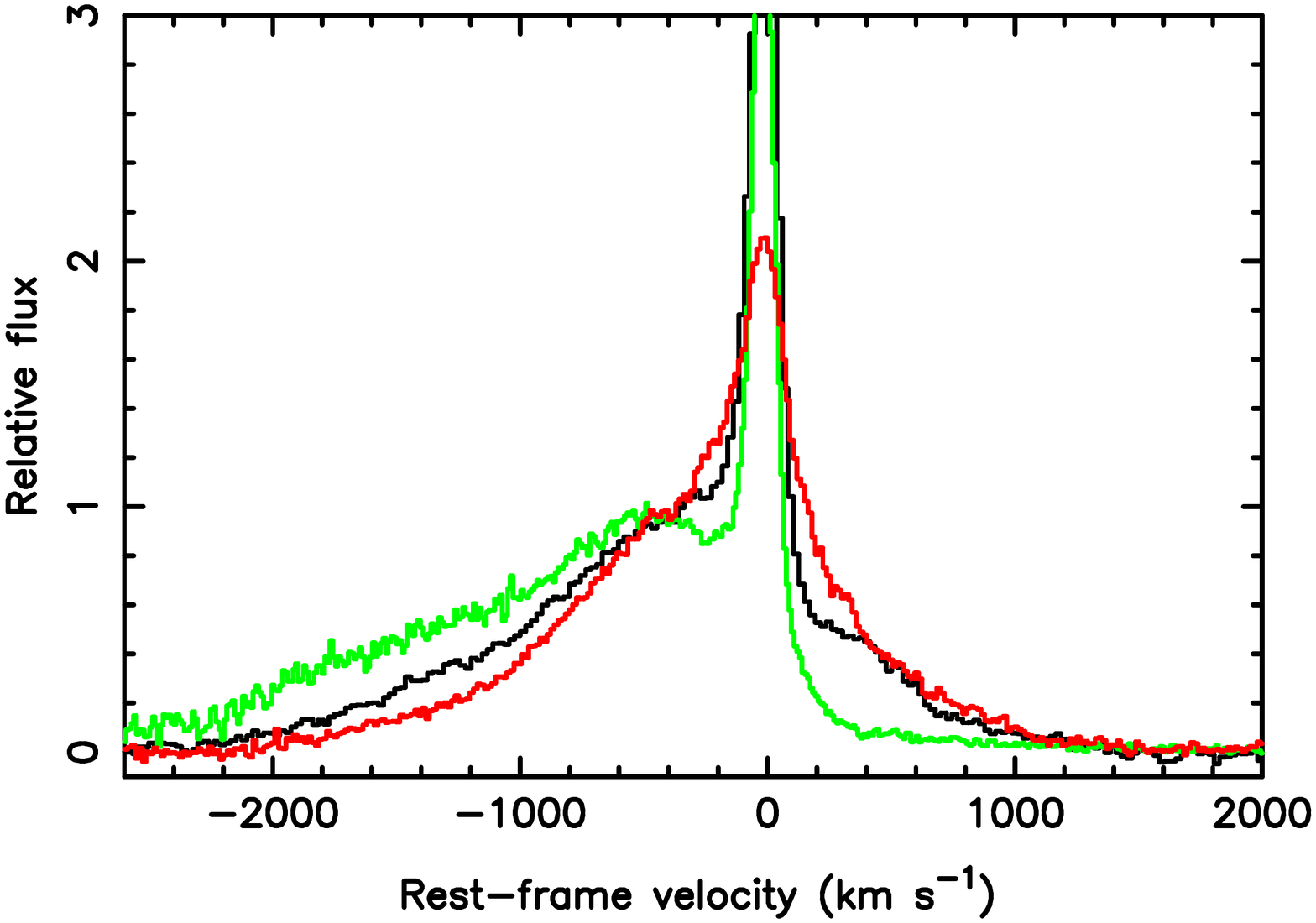}
\includegraphics[width=9.5cm]{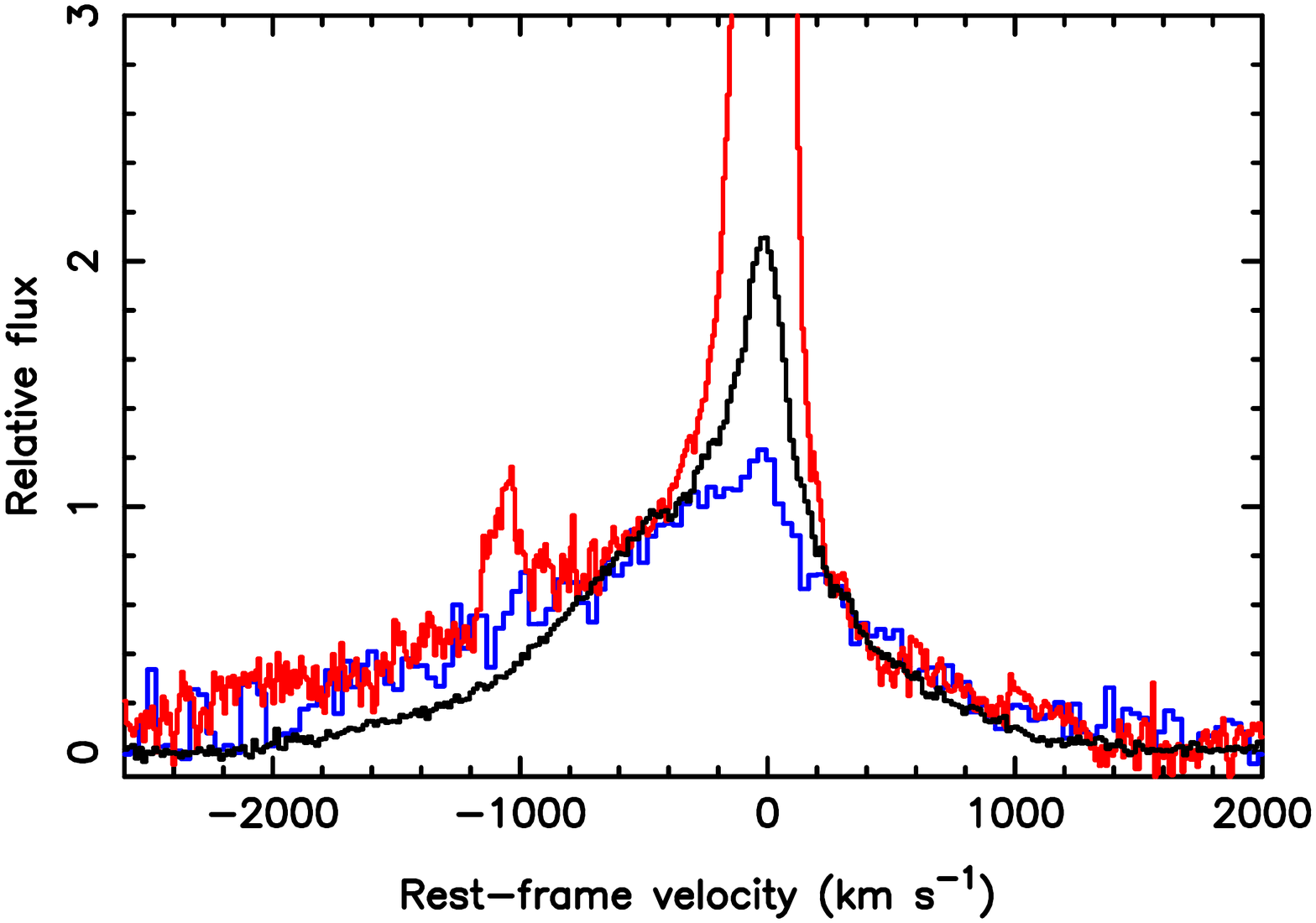}
\caption{Comparison of the velocity profiles of various emission lines detected in the 2018 Xshooter spectrum of F01004-2237. Top: profile comparison for [OIII]$\lambda$5007 (green), H$\beta$ (black) and HeI$\lambda$10830 (red). Bottom: profile comparisons for  OIII$\lambda$3133 (blue), HeI$\lambda$10830 (black), Pa$\alpha$ (red). Note that the Pa$\alpha$ velocity profile is contaminated by a HeI emission line at $\Delta V \sim -1000$\,km s$^{-1}$. For ease of comparison, all the profiles have been normalised to the flux measured at a velocity $\Delta V \sim -500$\,km s$^{-1}$.}
\label{fig:profiles}
\end{figure}

The case of H$\beta$ is more ambiguous:  the integrated flux of its broad/intermediate components appears to have not significantly varied between 2015 and 2018 (Table \ref{table:fluxes}). However, the profile comparison presented in Figure \ref{fig:blue} shows significant changes in the line profile: whereas the red wing of the line -- most likely related to the flare event -- has decreased in flux, the blue wing -- potentially containing a component related to the outflow detected in the forbidden lines -- has remained constant or slightly increased over some velocity ranges in the profile. Therefore, the relative constancy of the broad/intermediate H$\beta$ components is likely to be due to a combination of an increase in the flux of the outflow-related components, and a simultaneous decrease in the flux of the flare-related component. This interpretation is reinforced by consideration of the HeI$\lambda$5876 profile (see Figure \ref{fig:red}) for which the red wing has dramatically declined (by a larger factor than for H$\beta$), but the blue wing appears to have barely changed since 2015. Again, these trends can be explained in terms of a combination of a simultaneous decline in the flare-related component and increase in the outflow component, with the flare-related component making a larger contribution in the case of the HeI line than for H$\beta$ line. However, the composite nature of these permitted lines makes it challenging to unambiguously separate the contributions of flare- and outflow-related components, especially in the blue wings of the lines where the outflow components are likely to be strongest.

The composite nature of the profiles of the permitted lines also affects our ability to classify F01004-2237 in terms of whether it is a type I or a type II AGN. Certainly, the line ratios of the outflow-related components of the forbidden line are consistent with photoionisation of the broad/intermediate components of its forbidden lines by an AGN \citep{rz13}. However, the potential presence of strong broad wings related to the warm outflow (unambiguously detected in the forbidden lines), combined with the relatively low resolution of previous spectra, have made it challenging to identify broad wings to the permitted lines that might otherwise indicate a type I AGN. In this context, the comparison between the profiles of H$\beta$, HeI$\lambda$10830, [OIII]$\lambda$5007, OIII$\lambda$3133 and Pa$\alpha$ from our higher resolution Xshooter spectrum shown in Figure \ref{fig:profiles} is particularly illuminating. All the permitted lines, including the Bowen OIII$\lambda$3133 line, show a red wing extending to redshifted velocities $\ge$1000\,km s$^{-1}$ that is not present in the profile of the strong [OIII]$\lambda$5007 line. By itself, this is consistent with the presence of a type I AGN nucleus, although it does not entirely rule out TDE-related activity. Moreover, the relatively narrow half width at half maximum ($HWHM <$1000\,km s$^{-1}$) of the red wings of the permitted lines is consistent with a narrow-line Seyfert 1 (NLSy1) classification \citep[see][]{trakhtenbrot19}.

Clearly, the permitted lines differ in their detailed profiles (e.g. the red wing is weaker and the blue wing stronger in the H$\beta$ line than in the HeI$\lambda$10830 line), but this is likely to be at least partly due to varying relative contributions of the flare- and outflow-related components. It is also notable that the multiple Gaussian model that best fits the HeI$\lambda$10830 profile also provides a good fit to the  OIII$\lambda$3133 profile, albeit with different relative fluxes between the different kinematic components. Moreover, unconstrained multiple Gaussian fits to both of these lines require a very broad kinematic component ($FWHM \sim 2000$\,km s$^{-1}$) that is broader than found in any other forbidden or permitted line apart from HeII$\lambda$4686.

\section{Discussion}

\subsection{The nature of the warm outflow}

We now discuss the implications
of the forbidden line variability for our understanding of the warm outflows in F01004-2237.

\subsubsection{The compactness of the warm outflow region}

\begin{figure}
\includegraphics[width=7.5cm]{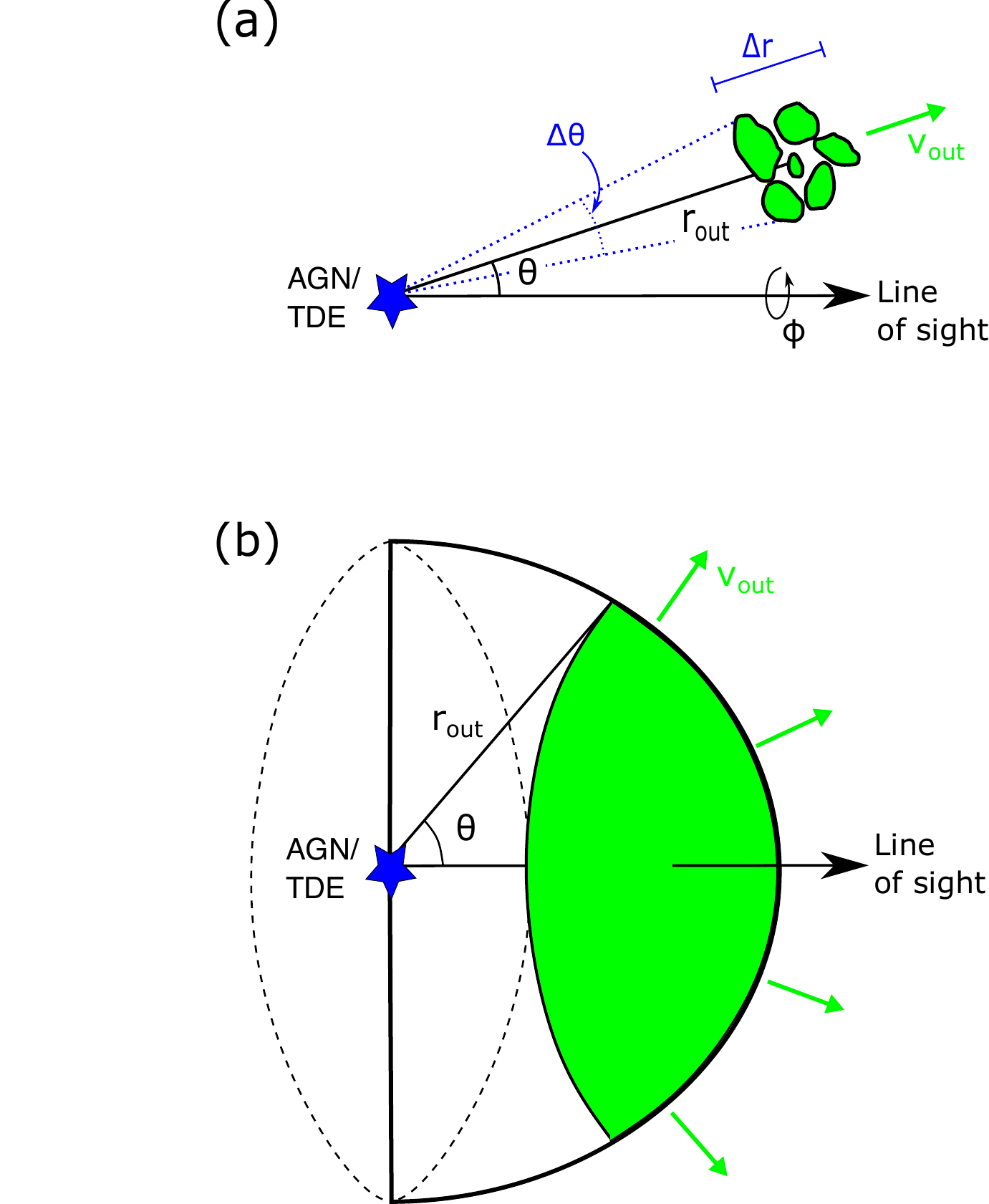}
\caption{Possible geometries for the outflow illuminated by the continuum flare: (a) a single cloud complex; and (b) a thin, hemispherical shell. In the latter case, the part of the illuminated outflow seen by the observer
is highlighted in green. Due to light-travel-time effects, $\theta$ will increase with time following the detection of the continuum flare by the observer. 
}
\label{fig:geometry}
\end{figure}

The most plausible explanation of the substantial increase in the broad/intermediate forbidden line fluxes between 2015 and 2018 is that  we are observing the effects of the enhanced ionisation of the warm outflow region due to its illumination by the 2010 flare.       
The time lag between the flare and the increase in the forbidden line flux -- in this case 5-8\,yr -- then provides information on the radial extent of the outflow region. For the simplest geometry of a single cloud or small complex of clouds (see Figure \ref{fig:geometry}(a)), the radius of the outflow region ($r_{\rm out}$) is related to the time lag ($\Delta t$) by:
\begin{equation}
r_{\rm out} = \frac{\Delta t c}{1-cos(\theta)}, 
\end{equation}
where $\theta$ is the angle between our line-of-sight and the radius vector from the nucleus to the outflow.
The major uncertainty in using the time lag to estimate the outflow radius is the unknown angle $\theta$. A lower limit is obtained by assuming that all the gas in the outflow is distributed perpendicular to the line-of-sight (i.e. $\theta=90$\,deg). In this case $1.5 < r_{\rm out} < 2.5$\,pc for $5 < \Delta t < 8$\,yr. However, in reality the outflow is unlikely to be distributed in this way, because the forbidden lines would not then show the substantial blueshifts that we measure for them. If we assume that the clouds in the outflows of ULIRGs are randomly distributed in the hemisphere on the side of the nucleus facing the observer, with the hemisphere on the other side of the nucleus obscured by dust,  then the most probable angle -- the solid-angle weighted mean -- is $\theta = 57.3$\,degrees  and we have $2.8 < r_{\rm out} < 4.6$\,pc.

The fact that there was apparently no significant increase in the forbidden-line flux by September 2015, but then the flux rose substantially some time between the September 2015 and August 2018, provides important information on the range of angles $\theta$ covered by the outflow and its radial thickness: $\Delta \theta$ and $\Delta r$ (see Figure \ref{fig:geometry}(a)).  Here we concentrate on determining  $\Delta r$ and  $\Delta \theta$ for the part of the outflow illuminated by the flare that was visible in 2018. Note, however, that we cannot entirely rule out the idea that emission from other spatial locations in the flare-illuminated outflow  might have been visible before this epoch, but by 2018 had substantially faded due to a short recombination time for the illuminated gas (see discussion in Sections 5.1.2 and 5.1.3 below).

Considering first the radial thickness, we start by assuming that the range in $\theta$ is negligible and that the forbidden-line flux rose to a maximum in 2018 -- our last observation epoch. In this case, we find that the ratio of the radial thickness to the maximum possible radial extent of the outflow ($r_{\rm max}$) is given by 
\begin{equation}
\frac{\Delta r}{r_{\rm max}} = \frac{8 - \Delta t_s({\rm yr})}{8}
\end{equation}
independent of the value of $\theta$, where $\Delta t_s$ is the lag between the time of the peak of the continuum flare and the time at which the forbidden line flux started to increase (i.e. $5 < \Delta t_s < 8$\,yr). We obtain a maximum $\Delta r / r_{\rm max} = 3/8$ for $\Delta t_s = 5$\,yr. This corresponds to  $\Delta r = 0.94$\,pc and $\Delta r = 1.7$\,pc for $\theta = 90$\,degrees  and $\theta = 57.3$\,degrees respectively. Note that these estimates would be smaller if, in reality, the forbidden-line fluxes from the outflow peaked before August 2018, or the outflow had a significant range in $\theta$. Therefore, we deduce that the outflow occupied a relatively thin layer in the radial direction.

In terms of the range in $\theta$, we start by considering the idealised geometry shown in Figure \ref{fig:geometry}(b), where the region of the outflow that has been illuminated by the flare is confined to uniformly filled, hemispherical shell of radius $r_{\rm out}$ on the observer's side of the nucleus that has a negligible radial thickness. Due to light-travel-time effects we see to larger angles $\theta$ around the shell as $\Delta t$ increases (see equation 1): initially,  only  the part of the hemisphere closest to the line of sight is visible, but eventually we see the whole hemisphere. In this case, the flux measured for the emission lines emitted by the shell will steadily rise to a maximum at time $\Delta t_{max} = r_{\rm out}/c$, when all of the illuminated hemispherical shell is visible to the observer (corresponding to $\theta = 90$\,degrees)\footnote{Note that such behaviour would be observed even if the emission from the part of the hemisphere closest to the observer had started to fade due to recombination effects before $\Delta t_{max}$, because the equatorial sectors of the hemisphere cover a larger area than the polar sectors close to the line of sight.}. However, in reality, we did not observe this steady increase in the forbidden-line flux for F01004-2237 between 2010 and 2018, but rather all the increase appears to have occurred after September 2015. This suggests that the region of the outflow associated with the flux change covered a
limited range in angle $\theta$.

\begin{figure}
\includegraphics[width=9.5cm]{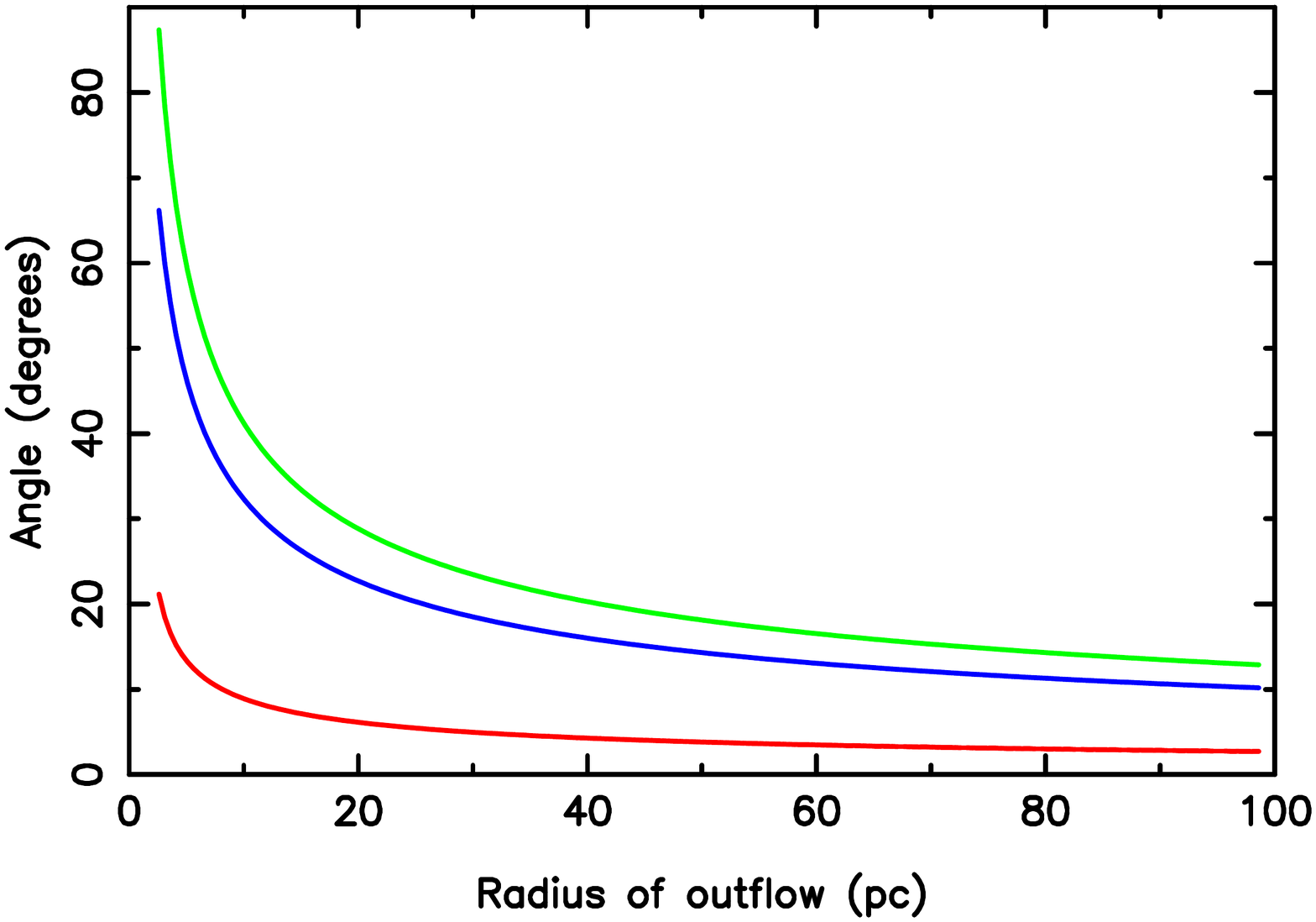}
\includegraphics[width=9.5cm]{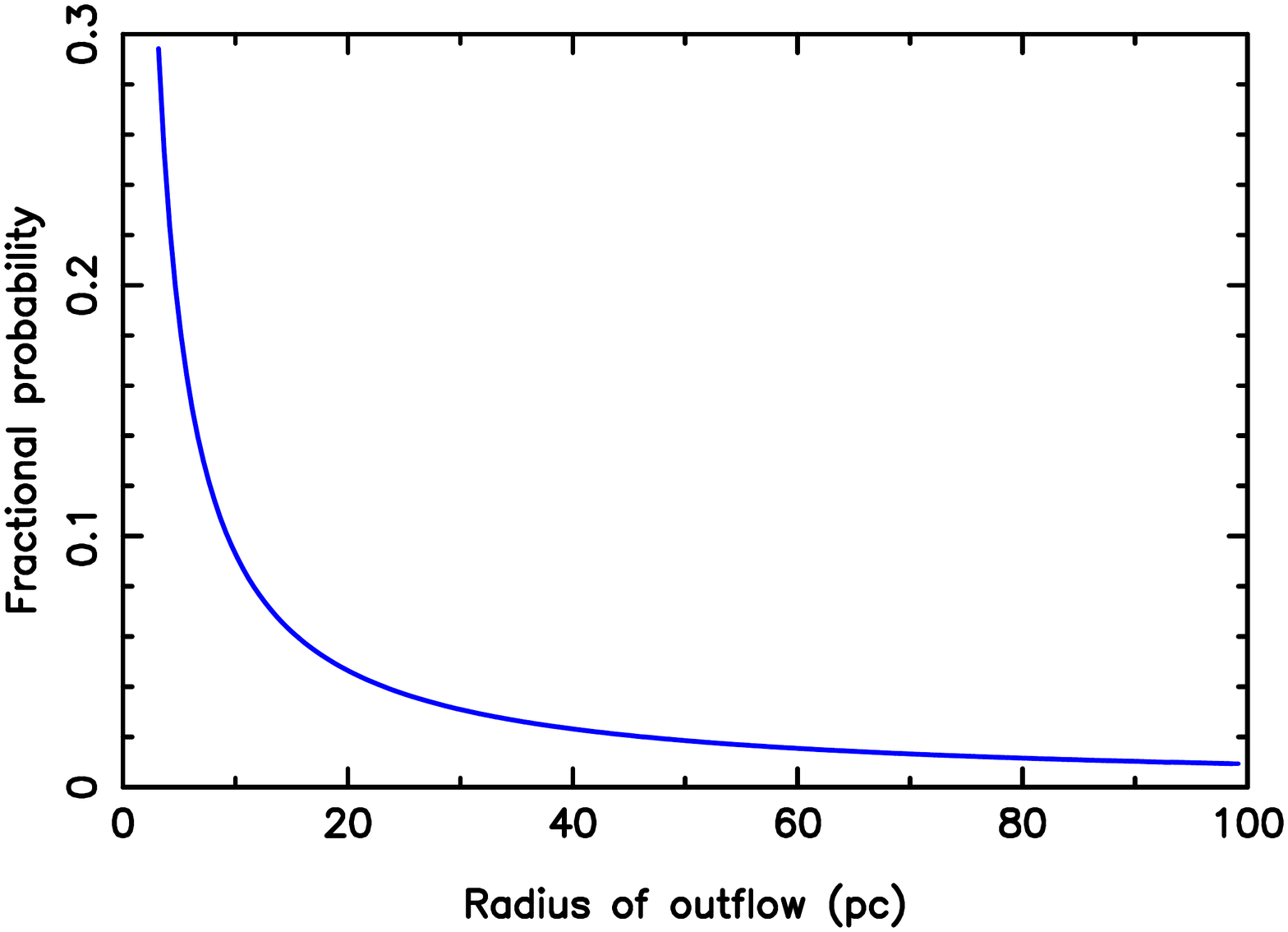}
\caption{Constraints on angle ($\theta$) relative to the line of sight and  outflow radius. Top: the angle between the nucleus--outflow vector and the line of sight ($\theta$) as a function of outflow radius $r_{\rm out}$ for time lags of $\Delta t = 8$ years (green) and $\Delta t = 5$ years (blue); and the difference between these two angles ($\Delta \theta$) for a given radius (red). Bottom: the fraction of the  hemisphere covered by the range of angles allowed by the time lags (from the top plot) as a function of $r_{\rm out}$. Assuming random angles ($\theta$) of the outflow relative to the line of sight, this is equivalent to the probability that the outflow has a particular $r_{\rm out}$.}
\label{fig:angles}
\end{figure}

We can estimate the range in angle ($\Delta \theta$)  by using equation 1 to determine how $\theta$ depends on $r_{\rm out}$ for particular time lags $\Delta t$. Figure \ref{fig:angles}~(top) shows the dependence of $\theta$ on $r_{\rm out}$ for the limiting minimum (5\,yr) and maximum (8\,yr) time lag for the increase in the [OIII] flux, and the difference between the angles ($\Delta \theta$) defined by the time 
lags. It is clear from these results that, regardless of $r_{\rm out}$, the region of flare-illuminated outflow visible in 2018 must have covered a relatively narrow range of angles: $\Delta \theta \sim 20$\,\,degrees for $r_{\rm out} \sim 3$\,pc and
$\Delta \theta < 5$\,\,degrees for $r_{\rm out} > 50$\,pc; the $\Delta \theta$ values would be even smaller if there were a range in $r_{\rm out}$. 
This implies that, if the illuminated outflow were at a relatively large distance from the nucleus (e.g. $r_{\rm out} > 50$\,pc), then it would be likely to comprise a small complex of clouds close to the line of sight (i.e. similar to the geometry shown in Figure \ref{fig:geometry}(a)). On the other hand, if it were situated at small nuclear distances (e.g. $r_{\rm out} < 10$pc), it could consist of a system of clouds with a wider angular extent (as seen from the nucleus), distributed closer to the plane of the sky -- perhaps material being ablated
from the circum-nuclear torus by a hot, radiation-driven wind.  Note, however, that these observations provide no information about the range in azimuthal angles $\phi$ around the line of sight (see Figure \ref{fig:geometry}(a)) covered by the outflow: for a given $\Delta \theta$ and $r_{\rm out}$ we cannot distinguish, for example, between a clump of clouds with small $\Delta \phi$ or
a  ring-like geometry with $\Delta \phi = 360$\,degrees.

Assuming that the outflows in such sources are distributed at random angles relative to the line of sight, we can use the information
in Figure \ref{fig:angles}~(top) to determine the most likely radial distance of the outflow from the nucleus. Figure \ref{fig:angles}~(bottom) shows the proportion of the hemisphere
covered by the range of angles ($\Delta \theta$) estimated for a given $r_{\rm out}$,\footnote{The proportion of the hemisphere covered by the range of angles $\Delta \theta$ encompassed by the outflow is given by $f_{H} = cos(\theta_5) -cos(\theta_8)$, where $\theta_5$ and $\theta_8$ are the angles calculated using equation 1 for time lags of 5 and 8\,yr respectively for a particular $r_{\rm out}$, and $\Delta \theta = \theta_8 - \theta_5$.} and hence the probability that the outflow is at a particular radial distance. This demonstrates that large outflow radii are unlikely ($<$2\% chance for $r >50$\,pc), thus  adding to the growing evidence that the warm outflow regions in nearby ULIRGs with optical AGN nuclei are spatially compact \citep{rose18,spence18,tadhunter18,tadhunter19}.

Overall, we deduce that part of the outflow that varied in flux is likely to have been situated relatively close to the illuminating AGN/TDE, had a small thickness in the radial direction, and covered a narrow range of angles $\theta$ as seen from the nucleus. Note that we have assumed that the source has not flared again since the last CSS observations at the end of 2015. If the variations in the forbidden lines represent the response of the gas in the outflow to a more recent (unseen) flare, then the outflow would need to be even more compact than we have estimated above, in order for the lines to show significant variability by 2018.

\subsubsection{Future evolution of the forbidden line fluxes}

It is also interesting to consider how the forbidden-line flux from the outflow is likely to evolve in the future. Assuming that the continuum flare had just encompassed the full radial extent of the outflow by August 2018, and that we could also see the full range of $\theta$ covered by the flare-illuminated outflow by that date, then the subsequent evolution of the forbidden-line emission will depend on a combination of two timescales: the time over which the continuum flare was bright ($\Delta t_f$), and the recombination timescale of the flare-illuminated gas in the outflow ($\Delta t_{rec}$). 

The timescale of the flare is hard to estimate precisely, because the CSS light curve presented in \citet{tadhunter17} is poorly sampled, and the time of the peak of the flare is only known to within  $\sim$6 months. However, $0.5 < \Delta t_f < 1.0$\,yr would be a reasonable estimate. 

An estimate of recombination timescale is given by $\Delta t_{rec} \sim 1/(\alpha_{\rm O^{2+}} n_e)$, where $\alpha_{\rm O^{2+}}$
is the recombination rate coefficient for the O$^{2+}$ ion, and $n_e$ is the electron density. In \citet{spence18} we deduced $n_e \sim 10^4$\,cm$^{-3}$ for the outflow in F01004-2237 from the ratios of the trans-auroral emission lines. Using this density estimate, along with the value of the O$^{2+}$ recombination coefficient given in \citet{osterbrock06} for a temperature $T = 10^4$\,K ($\alpha_{\rm O^{2+}} = 1.72\times10^{-11}$\,cm$^{-3}$ s$^{-1}$) we obtain $\Delta t_{rec} \sim 0.2$\,yr\footnote{Note that the recombination timescale for Fe$^{6+}$ is similar, assuming the same density and $T = 2\times10^4$\,K.}. We note that this is shorter than the estimated flare timescale ($\Delta t_f$). 

Therefore, if our assumptions are correct, we would expect the forbidden-line flux of the outflow component to have substantially faded on a timescale of one year or less following our last observations in 2018. However, if the flare had not in reality fully encompassed the full radial extent of outflow by August 2018 (but see Section 5.1.3 below) and/or light travel time effects meant that we could not see the full range of $\theta$ covered by the outflow at that epoch, then the fading timescale could be longer.

\subsubsection{Kinematics}

It is striking that we see evidence for variability over the entire blue wing of the strong [OIII]$\lambda$5007 line (Figure \ref{fig:blue}), including both the broad and intermediate kinematic components, rather than  being confined to discrete kinematic components covering narrow ranges of velocity. As well as suggesting that by 2018 the continuum flare had encompassed much of the outflow region\footnote{The alternative would be that the part of the outflow that we saw illuminated by the continuum flare in 2018 represents only a small fraction of total volume of the warm outflow. However, in that case the outflow as a whole would be required to show remarkable uniformity in its kinematic properties (e.g. velocity shifts and FWHM) at different spatial locations, otherwise it would be unlikely that the whole blue wing would vary in the way we observe.}, this result provides information on the nature of the emission-line kinematics. In particular, since we have demonstrated above that the range of angles ($\Delta \theta$) covered by the outflow is relatively small, it is unlikely that the large velocity range covered by the broad blue wing
($-2500 < v < -400$\,km s$^{-1}$) is due solely to different projections of the outflow velocity vector relative to the line of sight. In terms of projection effects, such a wide spread could only be explained if much of the outflow were moving in a direction very close to the plane of the sky. However, in that case the large overall projection factors would require the true (de-projected) outflow velocity to be much larger than the maximum velocity measured in the line profile. Therefore, it is more plausible that the warm gas in the outflow has a large {\it intrinsic} spread of velocities, perhaps as a result of the shredding of denser gas clouds by their hydrodynamic interaction with a faster, hotter wind of low density material \citep[e.g.][]{klein94}. This is in line with HST/STIS results for the spatially-resolved outflows in the ULIRG F14394+5332 \citep{tadhunter19}, the sample of nearby type II quasars studied by \citet{fischer18} and the archetypal Seyfert 2 galaxy NGC1068 \citep{das06}, which suggest that the local velocity dispersion of the warm gas is high across the full $\sim$1\,kpc extents of the outflows.


\subsubsection{Ionisation}

The fact that we see significant variability across emission lines with a range of ionisation, encompassing highly ionised species such as Ne$^{4+}$ and Fe$^{6+}$ as well as neutral species such as O$^{0}$, demonstrates that the emission lines of different ionisation in the outflow are produced at similar radial distances from the nucleus. The variability in the [OI]$\lambda$6300 line also provides evidence that some of the gas structures in the outflow have remained optically thick to the ionising continuum following illumination by the flare. 

It is also notable that the [NeV]$\lambda$3426 and [FeVII]$\lambda$6087 line fluxes showed the most pronounced increase in flux between 2015 and 2018. Concentrating on the integrated fluxes in the broad and intermediate kinematic components measured from the 2018 VLT/Xshooter spectrum, we find that these lines have the following ratios compared with the lower ionisation [NeIII]$\lambda$3869 and H$\beta$ lines: $[NeV](3426)/[NeIII](3869) = 1.32\pm0.18$ and $[FeVII](6087)/H\beta = 0.14\pm0.01$\footnote{Note that this is a lower limit on the $[FeVII](6087)/H\beta$ ratio for the warm outflow, since the broad and intermediate components of $H\beta$ are likely to have substantial contributions from the broad-line region associated with the flare.}. Based on AGN photoionisation models, such large ratios of the fluxes of high- to lower-ionization emission lines can be produced by a high ionisation parameter \citep[e.g.][]{taylor03}, an enhanced contribution from gas components that are optically thin to the ionising continuum \citep{binette96,binette97}, and/or an ionising spectral energy distribution (SED) that peaks at EUV wavelengths \citep[e.g.][]{moorwood96,ardila20}; some or all  of these conditions could result from the illumination of the outflow by the 2010 continuum flare. At the very least, the strengths of these high ionisation lines require the SED  of the flare to have a substantial flux of EUV photons with energies greater than the ionization energies of the Ne$^{3+}$ and Fe$^{5+}$ ions (i.e. $h\nu > 95$\,eV).

\subsection{The nature of the flaring activity}

\citet{trakhtenbrot19} have suggested that F010404-2237 and two other variable objects (AT2017bgt and OGLE17aaj) with similar properties, do not represent TDEs. Rather, they define a new class of flaring AGN, which is characterised by evidence for pre-existing AGN activity prior to the flare, slow decay times in the optical light curves following the flare compared with TDEs, and strong Bowen resonance-fluorescence lines of OIII and NIII.

In terms of this AGN interpretation for the flaring objects,  the strong Bowen lines suggest conditions that are different from typical AGN which lack such lines. The production of strong Bowen lines is favoured by an ionizing SED that peaks in the EUV, large optical depths in the line emitting region, and relatively high nitrogen abundances (to produce the NIII lines) \citep{netzer85,trakhtenbrot19}. As discussed in \citet{trakhtenbrot19}, the strength of the Bowen lines in these objects is most likely to be related to the flares having SEDs that are unusually strong at UV wavelengths. However,  the mechanism causing the flares remains uncertain in the AGN case. Therefore, it is not currently understood why the unusually UV-strong SEDs should be associated with these particular AGN flaring events,  but not with typical type I AGN or the majority of "changing look" AGN, which can show large increases in flux on a timescale of years, yet do not show the strong Bowen features. Moreover, the Bowen resonance-fluorescence lines have recently been detected in some flaring events that are typical of TDEs in terms of their light-curve decay characteristics \citep{blagorodnova19,onori19,leloudas19}. Therefore, the strong Bowen lines are not a unique indicator of AGN activity.

Apart from the strong Bowen lines, the HeII(4686)/H$\beta$ and HeI(5876)/H$\beta$ ratios for the integrated broad-line fluxes measured in the 2015 WHT/ISIS spectrum of F01004-2237 are also much higher than measured in the majority of type I AGN, but consistent with those measured in some TDEs.
The TDE explanation is also potentially supported by the fact that all three of the putative flaring AGN candidates have relatively low SMBH masses ($\sim10^{7}$\,M$_{\odot}$), since low black hole masses favour TDEs \citep[e.g.][]{wevers17}. 

The main problem with the TDE explanation is the long timescale of the post-flare decline in the activity. Although we lack photometric light-curve data for F01004-2237 from after the end of 2015, it is clear that some of the spectral features directly associated with the flare event were still present in 2018, approximately 8\,yr after the flare. In particular, the permitted lines (OIII$\lambda$3133, HeII$\lambda$4686, H$\beta$, HeI$\lambda\lambda$5876,10830, Pa$\alpha$) show broad, red wings in our Xshooter spectra that are not detected in the optical and infrared forbidden lines, albeit with reduced fluxes compared with our 2015 WHT/ISIS spectra. This long-timescale persistence of the broad spectral features is certainly not characteristic of most TDEs, which typically decay on timescales $<$1\,yr.

We speculate that, rather than being explained as {\it either} a TDE {\it or}  an AGN, these unusual events may represent {\it a combination} of the two phenomena: objects in which TDEs have occurred in galaxies with existing or dormant AGN activity. In this case, it is possible that the prolonged nature of the flare-related activity is due to the disruption of the AGN accretion disk by the debris stream of the TDE, subsequent re-settling of the AGN accretion disk, as well as re-processing of the light of the TDE flare by extended AGN-related gas structures such as the outer part of the accretion disk or the circum-nuclear torus. In this context, it is interesting that some ``changing-look'' AGN have also been explained in terms of hybrid AGN/TDE models \citep{merloni15,ricci20}. However, the latter cases do not show the strong Bowen lines visible in F01004-2237 and the two similar objects.

The modelling of the interactions between TDE tidal streams and AGN accretion disks
is in its infancy, but the simulations presented in \citet{chan19} clearly demonstrate that interactions with TDE debris have the potential to cause substantial disruption of AGN accretion disks, and lead to Eddington or super-Eddington accretion rates onto the super-massive black holes (SMBH). Although the initial continuum flare associated with these substantially enhanced accretion rates is likely to be short lived ($<< 1$\,yr), \citet{chan19} remark that, under certain circumstances, the accretion disk could take decades to settle to its unperturbed state. In this case, if the long-term settling of the accretion disk results in some enhancement in accretion rate onto the SMBH -- even if at a more modest level than that causing the initial flare -- then perhaps this could explain the long-lived nature of the outbursts in F01004-2237 and similar objects. 

\section{Conclusions}

We explain the late-time spectral evolution of the broad emission features in the UV/optical/near-IR spectrum of F01004-2237 in terms of the increasing contribution of a light echo from the warm outflow and a declining contribution from the direct emission of the flare-related components. The light echo -- observed in the near-UV and optical forbidden lines -- provides further evidence that the near-nuclear outflows in ULIRGs are compact, while the flare-related features -- observed in the broad, red wings of the permitted lines -- provide information on the nature of the flaring event. 

We confirm that some of the unusual features of the 2015 spectrum of the F01004-2237, such as broad NIII lines and the unusual profile of the [NeV]$\lambda$3426 lines, were produced by the Bowen resonance-fluorescence mechanism. This provides evidence that the flare had a continuum SED that extended to EUV, and illuminated material with a high optical depth and large nitrogen abundance. However, we cannot unambiguously distinguish between the TDE and AGN explanations for the origin of the flare based on our new Xshooter spectrum alone. Future, longer-term monitoring of F01004-2237 and similar objects, as well as extensive modelling of the interactions between TDE tidal streams and AGN accretion disks, should help to illuminate this issue.

\section*{Acknowledgements}
CT and JM acknowledge support from STFC.  Based on observations collected at the European Organisation for Astronomical Research in the Southern Hemisphere under ESO programme 099.A-0403, and  the WHT telescope operated on the island of La Palma by the Isaac Newton Group of Telescopes in the Spanish Observatorio del Roque de los Muchachos of the Instituto de Astrofísica de Canarias The authors acknowledge the data analysis facilities provided by the Starlink Project, which was run by CCLRC on behalf of PPARC. This research has made use of the NASA/IPAC Extragalactic Database (NED) which is operated by the Jet Propulsion Laboratory, California Institute of Technology, under contract with the National Aeronautics and Space Administration. 
 
\section*{Data availability}

The reduced data underlying this article will be shared on reasonable request to the corresponding author, and the publically-available raw data are available for the ESO archive.





\vglue 0.3cm\noindent








\bsp	
\label{lastpage}
\end{document}